\documentclass[twocolumn,showpacs,amsmath,amssymb,floatfix]{revtex4}
\usepackage{graphicx}
\usepackage{dcolumn}
\usepackage{bm}
\begin{document}
%
\title{Next-to-Leading Order QCD Corrections to Heavy Quark Correlations 
in Longitudinally Polarized Hadron-Hadron Collisions}
\author{Johann Riedl}
 \email{johann.riedl@physik.uni-r.de}
\author{Andreas Sch\"afer}%
 \email{andreas.schaefer@physik.uni-r.de}
 \affiliation{Institut f\"ur Theoretische Physik, Universit\"at Regensburg, 
              93040 Regensburg, Germany}
\author{Marco Stratmann}
 \email{marco@ribf.riken.jp}
\affiliation{Institut f\"ur Theoretische Physik, Universit\"at Regensburg, 
             93040 Regensburg, Germany\\
             Institut f\"ur Theoretische Physik und Astrophysik, 
             Universit\"{a}t W\"{u}rzburg, 97074 W\"{u}rzburg, Germany}
\begin{abstract}
We present a comprehensive phenomenological study of heavy flavor distributions and
correlations in longitudinally polarized proton-proton collisions at BNL-RHIC.
All results are obtained with a flexible parton-level Monte Carlo program at next-to-leading
order accuracy and include the fragmentation into heavy mesons, their subsequent semi-leptonic decays,
and experimental cuts. 
Next-to-leading order QCD corrections are found to be significant 
for both cross sections and double-spin asymmetries.
The sensitivity of heavy flavor measurements at BNL-RHIC to the gluon polarization of the nucleon
is assessed. Electron-muon and muon-muon correlations turn out to be the most promising
observables. 
Theoretical uncertainties are estimated by varying renormalization and factorization
scales, heavy quark masses, and fragmentation parameters.
\end{abstract}
\maketitle
%
\section{Motivation and Introduction}
%
Recent results from longitudinally polarized lepton-nucleon scattering
experiments \cite{ref:hermes,ref:smc,ref:compass-2had,ref:compass-charm}
and, in particular, for single-inclusive pion and jet production
in helicity-dependent proton-proton ($pp$) collisions at the 
Relativistic Heavy Ion Collider (RHIC) \cite{ref:phenix,ref:star} 
have started to put significant limits on the amount of 
gluon polarization in the nucleon \cite{ref:review,ref:dssv}.

This is best quantified in a ``global QCD analysis'', 
which treats all available experimental probes simultaneously and
consistently at a given order in the strong coupling 
$\alpha_s$ in perturbative QCD (pQCD). It allows one to extract the set of 
universal, spin-dependent parton distribution functions, defined as
\begin{equation}
\label{eq:pdfdef}
\Delta f(x,\mu) \equiv f_+(x,\mu) - f_-(x,\mu)\;,
\end{equation}
that yields the optimum theoretical description of the combined data.
In (\ref{eq:pdfdef}), $f_+$ ($f_-$) denotes the probability of finding
a parton of flavor $f=q,\bar{q},g$ at a resolution scale $\mu$ 
with light-cone momentum fraction $x$ and helicity $+$ ($-$) 
in a proton with helicity $+$.

Our current understanding of the spin structure of the nucleon
is derived from Eq.~(\ref{eq:pdfdef}) by taking the first moments
of the densities $\Delta f(x,\mu)$. These quantities enter 
the helicity sum rule of
the nucleon along with the contributions from the orbital angular
momenta of quarks and gluons \cite{ref:review}.
Specifically, the total gluon polarization is given by
\begin{equation}
\label{eq:firstmom}
\Delta g(\mu) \equiv \int_0^1 \Delta g(x,\mu) dx\;,
\end{equation}
and the challenge is to precisely map the gluon helicity density $\Delta g(x,\mu)$
in a wide range of $x$ in order to minimize extrapolation
uncertainties in the first moment $\Delta g(\mu)$.  

A first global QCD analysis of polarized parton densities $\Delta f(x,\mu)$
at next-to-leading order (NLO) accuracy was completed recently \cite{ref:dssv}.
It was based on the world-data on polarized inclusive  
and semi-inclusive deep-inelastic scattering, 
which are pivotal in constraining the quark and antiquark
densities \cite{ref:review}, as well as on the latest RHIC $pp$ measurements
\cite{ref:phenix,ref:star}  mentioned above.
The conclusion is that available results from all lepton-nucleon scattering experiments and 
the RHIC spin program are in nice agreement. 
This underpins the notion of factorization
also for spin-dependent hard scattering processes,
which is the foundation for most pQCD calculations and their predictive power.
The polarized gluon density $\Delta g(x,\mu)$ turns out to be compatible with zero in the
range of momentum fractions, $0.05\lesssim x\lesssim 0.2$, accessible to
experiments so far. However, it is still impossible to give a reliable estimate 
for the total gluon polarization $\Delta g(\mu)$ \cite{ref:dssv}. 
A significant contribution to the integral in Eq.~(\ref{eq:firstmom}) 
can still come from the unexplored small $x$ region. 
Hence, the fundamental question of what constitutes the proton spin 
still remains largely unanswered, despite the fact that impressive
progress, both theoretically and experimentally, was made 
in the past two decades. Back then, it was discovered
that only an unexpectedly 
small fraction, about a quarter, of the proton's spin 
can be attributed to the intrinsic spin of quarks and 
antiquarks \cite{ref:review}.

Narrowing down the uncertainties on $\Delta g(x,\mu)$
and, at the same time, extending the range in $x$ 
continues to be the main objective of experimental efforts in the years to come, 
utilizing both longitudinally polarized lepton-nucleon and proton-proton scattering. 
With higher luminosities becoming available at RHIC, less inclusive final-states
like jet-jet correlations will be instrumental in achieving this goal as they give a much
better handle on the $x$ range probed in experiment \cite{ref:spinplan}. 
Also, rare probes like prompt photons and heavy quarks come into focus. 
Both will not be able to compete with 
single-inclusive pion or jet measurements with respect to statistical precision, 
but they follow rather different underlying QCD hard scattering dynamics.
Therefore, such measurements are crucial for further testing and establishing the 
universality of helicity-dependent parton densities and hence for 
our understanding of the spin structure of the nucleon and QCD in general.

In this paper, we present a comprehensive phenomenological analysis of 
open heavy flavor production in longitudinally polarized $pp$ collisions at RHIC.
Until vertex detector upgrades are in place, RHIC experiments will identify
heavy quarks through their semi-leptonic decay electron or muon spectra, 
which receive contributions from both charm and bottom hadron decays, 
or by direct reconstructions of hadronic $D$ meson decays.
Since heavy flavors are a versatile probe of high-density medium effects in 
nucleus-nucleus collisions, such as modifications of the 
transverse momentum spectra \cite{ref:frawley},
various reference data have been taken at RHIC in unpolarized 
$pp$ collisions \cite{ref:phenix-hq,ref:star-hq}. 
Similar measurements are intended with longitudinally polarized protons \cite{ref:phenix-polhq}.

To reduce the uncertainties from deconvoluting experimental results 
for decay lepton spectra back to the heavy quark level, 
all theoretical calculations should be done as close as 
possible to the observational level. This was achieved, e.g., 
in a recent phenomenological study of unpolarized charm and bottom production 
at RHIC \cite{ref:cacciari}.
For the simplest example of a single-inclusive electron spectrum 
from semi-leptonic decays of a heavy meson $H_Q$, the corresponding 
invariant cross section takes schematically the following form
\begin{equation}
\label{eq:xsec-generic}
E_e\frac{d^3(\Delta)\sigma^e}{dp^3_e} = E_Q \frac{d^3(\Delta)\sigma^Q}{dp^3_Q} \otimes
D^{Q\rightarrow H_{Q}} \otimes f^{H_{Q}\rightarrow e}\;,
\end{equation}
where the symbol $\otimes$ denotes a convolution.
The cross section $d(\Delta)\sigma^Q$ for the production
of a heavy quark $Q$ with mass $m_Q$, energy $E_Q$, and 
momentum $p_Q$ in (polarized) $pp$ collisions
can be evaluated within pQCD. NLO QCD corrections,
which are essential for any meaningful, quantitative analysis,
are known, both in the unpolarized \cite{ref:nlo-unpol,ref:nlo-unpol2} and polarized  
\cite{ref:nlo-pol} case, for quite some time.
We note that the longitudinally polarized hadronic cross section
is defined as the combination
\begin{equation}
d\Delta\sigma^Q\equiv
\frac{1}{2}[d\sigma_{++}^Q-d\sigma_{+-}^Q] \,,
\label{eq:polxsec-def}
\end{equation}
where the subscripts $\pm$ label the helicity states of the
colliding hadrons.
The result for $d(\Delta)\sigma^Q$
depends on the choice of non-perturbative (helicity-dependent) parton densities,
the value of $m_Q$, and on the unphysical factorization ($\mu_f$) and 
renormalization ($\mu_r$) scales. The sensitivity of the cross section to
variations of $\mu_{f,r}$ can be taken as a rough estimate of the theoretical
uncertainty due to the truncation of the perturbative series at a certain order.
Likewise, variations of $m_Q$ contribute to the theoretical ambiguities
as well. We will assess all these sources of uncertainties 
in our detailed numerical studies.

The other two ingredients to Eq.~(\ref{eq:xsec-generic}),
are the fragmentation $D^{Q\rightarrow H_{Q}}$ 
of the heavy quark $Q$ into a heavy meson $H_Q$ and
the semi-leptonic decay $f^{H_{Q}\rightarrow e}$ of $H_Q$ into
the experimentally observed electrons $e$.
Since $m_Q$ cuts off final-state collinear singularities 
associated with the heavy quark, its hadronization $D^{Q\rightarrow H_{Q}}$
is fundamentally different from those for light quarks and gluons.
In the latter case, 
scale-dependent parton-to-hadron fragmentation functions \cite{ref:dss}
have to be introduced by virtue of the factorization theorem.
The non-perturbative transition $Q\rightarrow H_Q$ is described by
various phenomenological models for a scale independent function $D^{Q\rightarrow H_{Q}}$,
whose parameters are determined from fits to $e^+e^-$ data \cite{ref:frag-review}.
For our phenomenological studies, we use the functional form 
proposed in Ref.~\cite{ref:kart}, with its single parameter
taken in the range given in \cite{ref:frag-review}.
In addition, a fixed order pQCD calculation of $d(\Delta)\sigma^Q$ in 
Eq.~(\ref{eq:xsec-generic}) can be supplemented by all-order resummations of
quasi-collinear logarithms of the form $\alpha_s^n\log^n(p_T^Q/m_Q)$
\cite{ref:resum}, which can be large if the 
transverse momentum $p_T^Q$ of the produced
heavy quark is much larger than its mass.
For the time being, we do not pursue similar resummations for the
polarized hadroproduction of heavy quarks since
$p_T^Q\simeq m_Q$ for all phenomenologically relevant 
applications at RHIC.
We adopt the parameterization of the semi-leptonic decay spectrum 
$f^{H_{Q}\rightarrow e}$ obtained in Ref.~\cite{ref:cacciari}
from a fit to BaBar and CLEO data \cite{ref:babar-cleo}
and used in unpolarized analyses of heavy quark production at RHIC
\cite{ref:cacciari}.

The expression in Eq.~(\ref{eq:xsec-generic}) can be easily generalized
to the important case of heavy quark or decay lepton correlations.  
As will be demonstrated in some detail below, such measurements appear to
be more promising for accessing the gluon polarization at RHIC than 
single-inclusive decay electron or muon spectra.
To make theoretical calculations for such observables
feasible at NLO accuracy, we develop a flexible parton-level Monte Carlo program
to perform all phase-space integrations numerically.
For the subtraction of soft and collinear divergences present
at intermediate stages, we follow closely the methods devised in Ref.~\cite{ref:mnr} 
for the computation of heavy flavor correlations in unpolarized hadron-hadron collisions.

Our Monte Carlo code is capable of computing any infrared safe 
heavy flavor cross section at ${\cal{O}}(\alpha_s^3)$, including
correlations of the $Q\bar{Q}$ pair and control of the 
accompanying jet, with the same kinematic cuts as used 
in experiment.
The hadronization of the $Q\bar{Q}$ pair into heavy mesons 
and their subsequent semi-leptonic decays are modeled as outlined above.
Our results complement and significantly extend 
previous spin-dependent NLO calculations
of single-inclusive heavy (anti)quark yields
and of the heavy quark charge asymmetry
based on largely analytical methods \cite{ref:nlo-pol,ref:hans},
where any information on the partonic recoil system was lost.
The required spin-dependent matrix elements squared at ${\cal{O}}(\alpha_s^3)$
for producing a $Q\bar{Q}$ pair plus a light parton, 
\begin{equation}
\label{eq:nlo-proc}
gg \rightarrow Q\bar{Q}g\,\,\,,\,\,\,
q\bar{q} \rightarrow Q\bar{Q}g\,\,\,,\,\,\,
gq(\bar{q}) \rightarrow Q\bar{Q}q(\bar{q})\,,
\end{equation}
as well as the
virtual corrections to the ${\cal{O}}(\alpha_s^2)$ Born cross sections,
\begin{equation}
\label{eq:born-proc}
gg \rightarrow Q\bar{Q}\,\,\,,\,\,\,
q\bar{q}\rightarrow Q\bar{Q}\,,
\end{equation}
can be taken from Ref.~\cite{ref:nlo-pol}. 
Expressions obtained in $d=4-2\varepsilon$ dimensional regularization are
required only in the singular regions of phase-space, and $\varepsilon$ can be
set to zero otherwise.

After appropriate modifications, the results obtained in this work 
can be used also as the ``resolved'' photon contribution to the 
spin-dependent photoproduction of heavy quarks at NLO.
A similar parton-level Monte Carlo program including both direct and resolved
photon processes will be presented in a forthcoming publication \cite{ref:future}.
It will allow one to include existing data on spin-dependent charm photoproduction 
\cite{ref:compass-charm} into future global QCD analyses of polarized parton densities.
NLO expressions for the point-like, ``direct'' photon part of the
cross section can be adapted from \cite{ref:ingo-photo}; see also \cite{ref:polphoto2}. 

The outline of the paper is as follows: in Sec.\ II we briefly 
review some of the technical aspects of setting up a parton-level Monte Carlo program
for heavy flavor production in polarized hadron-hadron collisions at NLO accuracy.
Some additional technical details can be found in the Appendix.
In Sec.\ III we present a detailed phenomenological study of
heavy flavor production in polarized $pp$ collisions at RHIC, focussing on
experimentally relevant decay lepton, electron or muon, distributions and 
correlations within the kinematic acceptance of the 
PHENIX and STAR experiments at RHIC.
We assess theoretical uncertainties due to variations of $\mu_f$ and $\mu_r$,
the heavy quark mass $m_Q$, and parameters related to modeling the fragmentation
process. Finally, we give predictions for double-spin asymmetries and discuss their
potential in further constraining helicity-dependent parton densities, in particular,
$\Delta g(x,\mu)$.
We summarize our results in Sec.\ IV.

\section{Technical Framework\label{sec:tech}}
%
We consider heavy quark hadroproduction in longitudinally polarized 
$pp$ collisions at ${\cal{O}}(\alpha_s^3)$ in QCD. 
All phase-space integrations are performed numerically with Monte Carlo
techniques. This enables us to compute any observable involving heavy quarks
within experimental acceptance cuts, including single-inclusive distributions, 
correlations among the heavy quark-antiquark pair, and, although not pursued in this work, 
with the associated hard jet present for the first time at ${\cal{O}}(\alpha_s^3)$. 
This significantly extends available calculations \cite{ref:nlo-pol,ref:hans} 
based on largely analytical integrations over the variables characterizing the 
partons recoiling from the observed single-inclusive heavy quark or antiquark.

In the integration of the fully exclusive partonic cross sections 
for the processes in Eqs.~(\ref{eq:nlo-proc}) and (\ref{eq:born-proc})
at ${\cal{O}}(\alpha_s^3)$, one has to deal with ultraviolet, infrared,
and collinear divergences, which have to be eliminated before any numerical
approach can be used.
To this end, we follow closely the subtraction method devised 
and used in Ref.~\cite{ref:mnr} to compute 
the unpolarized production of a $Q\bar{Q}$ pair.
The subtraction method is based on adding and subtracting counter terms
which approximate the real emission processes in (\ref{eq:nlo-proc}) 
in the singular regions of phase-space and are integrable 
with respect to the momentum of an unresolved parton.
In this Section, we briefly review the technical aspects relevant to extend and apply the 
subtraction method of \cite{ref:mnr} to heavy quark hadroproduction 
in longitudinally polarized $pp$ collisions and discuss the numerical implementation.
For further details, we refer the reader to Ref.~\cite{ref:mnr}.
We note that a general formulation of the dipole subtraction method 
for NLO calculations with massive partons in QCD and supersymmetric QCD 
has been developed in \cite{ref:dipole}.

Assuming, as usual, factorization, the cross section
(\ref{eq:polxsec-def}) for producing a heavy $Q\bar{Q}$ pair in 
longitudinally polarized $pp$ collisions at a 
center-of-mass system (c.m.s.) energy $\sqrt{S}$ 
can be written as a convolution, 
\begin{eqnarray}
\label{eq:xsec-fact}
\nonumber
d\Delta\sigma^Q &=& \sum_{a,b}\int dx_1 dx_2 \Delta f_a(x_1,\mu_f) \Delta f_b(x_2,\mu_f)\,{\cal{S}}\\
&\times& d\Delta\hat{\sigma}_{ab}(x_1,x_2,S,m_Q,k_1,k_2,\mu_f,\mu_r)\,,
\end{eqnarray}
where the $\Delta f_{a,b}(x_{a,b},\mu_f)$ denote the spin-dependent 
parton distribution functions of flavor $a,b$ at momentum fraction $x_{a,b}$
and scale $\mu_f$, as defined in Eq.~(\ref{eq:pdfdef}).
The sum in (\ref{eq:xsec-fact}) is over all contributing partonic processes
$ab\rightarrow Q\bar{Q}c$ to ${\cal{O}}(\alpha_s^3)$
with $d\Delta\hat{\sigma}_{ab}$ the
associated polarized hard scattering cross sections. They are defined in complete
analogy to Eq.~(\ref{eq:polxsec-def}) and can be computed perturbatively as a series
in the strong coupling $\alpha_s$. Parton $c$ is either a gluon or a light
(anti)quark producing the associated jet possible at ${\cal{O}}(\alpha_s^3)$.
$k_{1,2}$ denote the momenta of the heavy quark $Q$ and antiquark $\bar{Q}$ 
with mass $m_Q$, i.e., $k_{1,2}^2=m_Q^2$. 
 
The required spin-dependent matrix elements squared at  
${\cal{O}}(\alpha_s^3)$ in $d=4-2\varepsilon$ 
dimensional regularization for the processes in 
(\ref{eq:nlo-proc}) and (\ref{eq:born-proc}) can be taken from
Ref.~\cite{ref:nlo-pol}.
Starting from the NLO level, the subprocess cross sections
$d\Delta\hat{\sigma}_{ab}$ in (\ref{eq:xsec-fact})
depend explicitly on the
renormalization and factorization scale $\mu_r$ and $\mu_f$,
arising from the subtraction of ultraviolet and collinear singularities,
respectively. Infrared (soft gluon) divergences cancel among real emission
and virtual loop corrections.

In Eq.~(\ref{eq:xsec-fact}), ${\cal{S}}$ is the ``measurement function''
used to define the observable one is interested in. One can think of
${\cal{S}}$ as being a set of step functions implementing
the experimental cuts imposed on the final-state particles
and selecting a certain bin in a histogram.
As mentioned in the Introduction, charm and bottom quarks 
are currently detected only indirectly at RHIC, mainly through the
semi-leptonic decays of the produced heavy $D$ and $B$ mesons.
Thus, the cross section (\ref{eq:xsec-fact}) at the heavy quark-level is
not yet sufficient for comparing theory with experimental results.
As indicated in Eq.~(\ref{eq:xsec-generic}), one needs to
convolute the parton-level results for $d\Delta\sigma^Q$
with additional phenomenological functions
$D^{Q\rightarrow H_Q}$ and $f^{H_Q\rightarrow e}$
describing the hadronization into a heavy meson $H_Q$ and
the semi-leptonic decay of $H_Q$ into the observed lepton,
respectively.
Our flexible parton-level Monte Carlo program not only performs the
phase-space integrations for arbitrary ${\cal{S}}$ 
for any infrared safe observable but can also
account for the semi-leptonic decays of the heavy quark pair
into electrons and muons.
We specify our choice for $D^{Q\rightarrow H_Q}$ and $f^{H_Q\rightarrow e}$
in Sec.~\ref{sec:sec3-a}.

For the implementation of Eq.~(\ref{eq:xsec-fact}) in a numerically efficient
integration it is convenient to express the three-body phase-space
and the matrix elements squared for the $2\rightarrow 3$
processes listed in (\ref{eq:nlo-proc})
in terms of variables in which soft and collinear singularities 
can be identified easily.
Instead of choosing the usual set of five independent scalar products 
(or Mandelstam variables) of the parton momenta in $ab\rightarrow Q\bar{Q}c$,
this is achieved by introducing $x$, $y$, $\theta_1$, $\theta_2$, and 
$s=x_1x_2S$ \cite{ref:mnr}. They are defined as follows:
$x=(k_1+k_2)^2/s$, the invariant mass of the
$Q\bar{Q}$ pair scaled by the available partonic c.m.s.\ energy squared, 
i.e., $\rho\equiv4m_Q^2/s\le x\le 1$,
and $y$ is the cosine of the angle between the $z$-direction,
aligned with the spatial direction of parton $a$, and $\vec{k}_3$, 
the momentum of parton $c$, in the c.m.s.~of the incoming partons, i.e., $-1\le y\le 1$.
Soft and collinear regions of phase-space are associated with
$x=1$ and $y=\pm 1$, respectively.
Both $\theta_1$ and $\theta_2$ do not matter for this discussion.
They range between $0$ and $\pi$ and are 
used to parameterize the spatial orientation of $k_{1,2}$ with respect
to the plane span by the other three momenta
in the c.m.s.\ of the $Q\bar{Q}$ pair, see \cite{ref:mnr} 
for an explicit parameterization of the momenta.

The $d$-dimensional three-body phase-space expressed in terms of the variables
$x$, $y$, $\theta_1$, $\theta_2$, and $s$ reads
\begin{eqnarray}
\label{eq:dps3}
\nonumber
d\mathrm{PS}_3 &=&\frac{1}{\Gamma(1-2\varepsilon)}2^{-9+6\varepsilon} \pi^{-4+2\varepsilon}
s^{1-2\varepsilon}\beta_x^{1-2\varepsilon} \\
\nonumber
&\times& x^{-\varepsilon}(1-x)^{1-2\varepsilon} dx\, (1-y^2)^{-\varepsilon} dy \\
&\times&\sin^{1-2\varepsilon}\theta_1 \, d\theta_1 \sin^{-2\varepsilon}\theta_2 \, d\theta_2\,,
\end{eqnarray}
which agrees with the result in \cite{ref:mnr} and where we have
introduced $\beta_x=[1-4m_Q^2/(sx)]^{1/2}$.
$\Gamma(z)$ represents the Gamma function.

The contribution of the $2\rightarrow 3$ real emission processes 
in (\ref{eq:nlo-proc}) is then given by
\begin{equation}
\label{eq:m-ps3}
d\Delta\hat{\sigma}_{ab}=
\Delta|M_{ab}|^2\, d\mathrm{PS}_3\,,
\end{equation}
where the spin-dependent amplitude squared, $\Delta|M_{ab}|^2$,
includes the partonic flux factor $1/(2s)$ and
is summed over final-state color and spin degrees 
of freedom and averaged over the color of the interacting partons
$a,b$ \cite{ref:nlo-pol}.
Soft ($x=1$) and collinear ($y=\pm1$) singularities in $\Delta|M_{ab}|^2$
appear as
\begin{equation}
\label{eq:m-sing}
\Delta|M_{ab}|^2=\frac{\Delta f_{ab}(s,m_Q,x,y,\theta_1\theta_2)}{s^2(1-x)^2(1-y^2)}\,,
\end{equation}  
where $\Delta f_{ab}$ is regular for $x=1$ and $y=\pm1$.
The $qg$ process in (\ref{eq:nlo-proc}) can have only collinear singularities at NLO.
Due to the finite mass $m_Q$, there can be no collinear gluon radiation from a 
heavy quark (``dead cone'').

Upon inserting (\ref{eq:dps3}) and (\ref{eq:m-sing}) into (\ref{eq:m-ps3}), one can proceed by
expanding the resulting $(1-x)^{-1-2\varepsilon}$ and $(1-y^2)^{-1-\varepsilon}$
for small $\varepsilon$ as shown in Ref.~\cite{ref:mnr}, 
\begin{eqnarray}
\label{eq:xy-expand}
\nonumber
(1-x)^{-1-2\varepsilon} &=& -\frac{\tilde{\beta}^{-4\varepsilon}}{2\varepsilon}\delta(1-x)+
\left( \frac{1}{1-x} \right)_{\tilde{\rho}}\\
\nonumber
&-&2\varepsilon \left(\frac{\log(1-x)}{1-x}\right)_{\tilde{\rho}} + {\cal{O}}(\varepsilon^2)\,,\\
\nonumber
(1-y^2)^{-1-\varepsilon}&=& -[\delta(1+y)+\delta(1-y)] 
\frac{(2\omega)^{-\varepsilon}}{2\varepsilon} \\
\nonumber
&+&\frac{1}{2} \left[ \left( \frac{1}{1-y} \right)_{\omega}+
\left( \frac{1}{1+y} \right)_{\omega} \right]+ {\cal{O}}(\varepsilon)\,,\\
\end{eqnarray}
where $\tilde{\beta}=\sqrt{1-\tilde{\rho}}$.
Explicit expressions for the
distributions $[1/(1-x)]_{\tilde{\rho}},\ldots$ in (\ref{eq:xy-expand}) 
are collected in Eq.~(\ref{eq:distrib}) of the Appendix.
The choice of the parameters $\tilde{\rho}$ and $\omega$ 
is to some extent arbitrary and will be discussed at the end of this Section.
Using Eq.~(\ref{eq:xy-expand}), the subprocess cross sections for $ab\rightarrow Q\bar{Q}c$ 
at ${\cal{O}}(\alpha_s^3)$ can be decomposed as
\begin{eqnarray}
\label{eq:xsec-decomp}
\nonumber
d\Delta\hat{\sigma}_{ab} &=&
d\Delta\hat\sigma^{(b)}_{ab}
+d\Delta\hat{\sigma}^{(c+)}_{ab}+d\Delta\hat{\sigma}^{(c-)}_{ab}\\
&+&d\Delta\hat{\sigma}^{(s)}_{ab}+d\Delta\hat{\sigma}^{(v)}_{ab}+d\Delta\hat{\sigma}^{(f)}_{ab}.
\end{eqnarray}
Here, $d\Delta\hat\sigma^{(b)}_{ab}$ and $d\Delta\hat{\sigma}^{(v)}_{ab}$
denote the ${\cal{O}}(\alpha_s^2)$ Born contribution and the
${\cal{O}}(\alpha_s^3)$ one-loop corrections to the 
$gg$ and $q\bar{q}$ scattering processes in (\ref{eq:born-proc}), respectively. 
Analytic expressions for the virtual 
contributions in $d$ dimensions, with ultraviolet divergences being subtracted at 
a renormalization scale $\mu_r$, have been obtained in Ref.~\cite{ref:nlo-pol}.

In Eq.~(\ref{eq:xsec-decomp}), $d\Delta\hat{\sigma}^{({s})}_{ab}$ is the soft component 
of the $gg$ or $q\bar{q}$ scattering cross section, which can
be either evaluated by explicitly taking the soft gluon limit of the full $d$-dimensional 
matrix elements squared computed in Ref.~\cite{ref:nlo-pol}
or constructed using general properties of 
soft gluon emission in QCD, see, e.g., \cite{ref:mnr}.
In the limit $x\rightarrow 1$, the kinematics simplifies, and phase-space 
integrations can be performed analytically.
The relevant integrals are the same as for unpolarized heavy flavor hadroproduction and
can be found, e.g., in App.~A of Ref.~\cite{ref:mnr}.
One obtains
\begin{eqnarray}
\label{eq:xsec-soft}
\nonumber
d\Delta\hat{\sigma}^{(s)}_{ab}&=&-\frac{\Gamma(1-\varepsilon)}{\Gamma(1-2\varepsilon)}
(4\pi)^{\varepsilon-3} s^{-1-\varepsilon} \tilde{\beta}^{-4\varepsilon}\\
&\times&\frac{1}{\varepsilon} \Delta f_{ab}^{(s)}(s,m_Q,\theta_1) d\mathrm{PS}_2\,.
\end{eqnarray}
Explicit expressions for 
$d\Delta\hat{\sigma}^{(s)}_{ab}$ and, for completeness, the  standard two-body
phase-space factor $d\mathrm{PS}_2$ in $d$ dimensions are given in the
Appendix.

All $2\rightarrow 3$ processes in (\ref{eq:nlo-proc}) 
exhibit singularities related to collinear
splittings off the incoming partons. Again, for such configurations 
the kinematics collapses to the simpler case of $2\rightarrow 2$ scattering,
and these contributions, summarized by $d\Delta\hat{\sigma}^{(c\pm)}_{ab}$
in Eq.~(\ref{eq:xsec-decomp}), can be evaluated analytically.
As for the soft contribution, one can either start by taking the collinear
$(y\rightarrow \pm 1)$ limit of the full, $d$-dimensional $2\rightarrow 3$ 
matrix elements taken from Ref.~\cite{ref:nlo-pol} or by deriving the
expressions from scratch. 
After combining the relevant matrix elements with $d{\mathrm{PS}}_3$ in Eq.~(\ref{eq:dps3}),
taking the limit $y\rightarrow \pm 1$, and integrating over $\theta_2$ 
one obtains
\begin{eqnarray}
\label{eq:collinear}
\nonumber
d\Delta\hat{\sigma}^{(c\pm)}_{ab} &=& -(4\pi)^{\varepsilon-2} \Gamma[1+\varepsilon] 
\left(\frac{2}{\omega}\right)^{\varepsilon} \frac{s^{-1-\varepsilon}}{4\varepsilon} d{\mathrm{PS}}_2^x
\\
\nonumber
&\times& \left[\left( \frac{1}{1-x} \right)_{\tilde{\rho}} -2\varepsilon 
\left( \frac{\log(1-x)}{1-x} \right)_{\tilde{\rho}} \right] \\
&\times& \Delta f_{ab}^{(c\pm)}(s,m_Q,x,\theta_1)\,,
\end{eqnarray}
where $dPS_2^x=dPS_2\big|_{s\rightarrow xs} dx$.
The superscript $\pm$ in $d\Delta\hat{\sigma}^{(c\pm)}_{ab}$ distinguishes
the two configurations with $y=+1$ and $y=-1$, where
parton $c$ is emitted collinearly to the momentum of 
parton $a$ and $b$, respectively. 
The relevant $\Delta f_{ab}^{(c\pm)}$ are again collected in the Appendix.

The last term in Eq.~(\ref{eq:xsec-decomp}), $d\Delta\hat{\sigma}^{(f)}_{ab}$,
contains all the finite contributions after using the expansions (\ref{eq:xy-expand})
for $(1-x)^{-1-2\varepsilon}$ and $(1-y^2)^{-1-\varepsilon}$,
and the phase-space integration
can be performed numerically in four dimensions, i.e., with $\varepsilon\rightarrow 0$.
One obtains
\begin{eqnarray}
\nonumber
\label{eq:xsec-finite}
d\Delta\hat{\sigma}^{(f)}_{ab}&=&\frac{1}{2^{10}\pi^4s} 
\left( \frac{1}{1-x}\right)_{\tilde{\rho}} \left[
\left(\frac{1}{1-y}\right)_{\omega} + 
\left( \frac{1}{1+y} \right)_{\omega} \right]\\
\nonumber
&\times&
\beta_x \sin\theta_1 d\theta_1\, d\theta_2\, dx\, dy\\
&\times& 
\Delta f_{ab}(s,m_Q,x,y,\theta_1,\theta_2)\,.
\end{eqnarray}
As can be seen, all soft and collinear 
singularities are regulated by the $\tilde{\rho}$- and $\omega$-prescriptions
defined in (\ref{eq:distrib}).

The resulting $1/\varepsilon$ divergence in (\ref{eq:collinear})  
assumes the form dictated by the factorization theorem, i.e., a
convolution of $d$-dimensional helicity-dependent LO splitting functions
$\Delta P_{ij}(x)$ and Born matrix elements $\Delta|M_{ab}|^2$.
Due to the collinear splitting,  
the latter have to be evaluated at a ``shifted kinematics'' where
parton $a$ (or $b$) carries only a fraction $x$ of its original momentum, 
i.e., $s\rightarrow xs$ and $d\mathrm{PS}_2 \rightarrow d\mathrm{PS}_2^x$;
see Eqs.~(\ref{eq:coll1})-(\ref{eq:coll6}) in the Appendix,
where, for convenience, also the Born cross sections and the
LO $\Delta P_{ij}(x)$ are listed.
Collinear singularities are factorized 
into the bare parton distribution functions at a scale $\mu_f$
by adding an appropriate ``counter cross section'' to (\ref{eq:collinear}) 
which to ${\cal{O}}(\alpha_s^3)$ schematically reads
\begin{eqnarray}
\label{eq:xsec-counter}
\nonumber
d\Delta\hat{\sigma}_{ab}^{\tilde{c}}(\mu_f)
&=&-\frac{\alpha_s}{2\pi} \sum_i \int \frac{dx}{x} 
\Big[ \Delta {\cal{P}}_{ia}(x,\mu_f) d\Delta\hat{\sigma}_{ib}^{(b)}(xs)\\
&+& \Delta {\cal{P}}_{ib}(x,\mu_f) d\Delta\hat{\sigma}_{ai}^{(b)}(xs)\Big]\,,
\end{eqnarray}
where
\begin{equation}
\label{eq:trans-func}
{\cal{P}}_{ij}(x,\mu_f) = \Delta P_{ij}(x) [-\frac{1}{\varepsilon}+\gamma_E-\ln 4\pi 
+\ln\frac{\mu_f^2}{\mu^2}] + \Delta g_{ij}(x).
\end{equation}
The sum in (\ref{eq:xsec-counter}) is over all possible collinear configurations
involving one of the initial-state partons $a,b$.
The argument $xs$ of the Born cross sections $d\Delta\hat{\sigma}_{ai}^{(b)}$ 
in (\ref{eq:xsec-counter}) indicates
that they have to be evaluated at the shifted kinematics as discussed above.
In Eqs.~(\ref{eq:xsec-counter}) and (\ref{eq:trans-func}), the Euler constant 
$\gamma_E$ and $\ln 4\pi$, both, like the scale $\mu$, 
artifacts of dimensional regularization,
are subtracted along with the $1/\varepsilon$ 
singularity. The factorization scheme is fully determined by the choice of $\Delta g_{ij}$,
for which we take $\Delta g_{qq}=-4C_F (1-x)$ 
with $C_F=4/3$ and $\Delta g_{ij}=0$ otherwise.
This guarantees helicity conservation when the HVBM prescription for $\gamma_5$ in 
$d$ dimensions is adopted to project onto definite helicity states \cite{ref:hvbm} and
defines the $\overline{\mathrm{MS}}$ scheme in the polarized case \cite{ref:nlo-split},
which we use throughout our calculations.
As a consequence of factorization, both the hard scattering cross sections and
the parton distribution functions in Eq.~(\ref{eq:xsec-fact}) 
depend on the scale $\mu_f$ which is arbitrary.
$\mu_f$ can be chosen differently than the renormalization scale $\mu_r$ at which 
ultraviolet singularities are absorbed into the bare coupling and heavy quark mass.

Note that $d\Delta\hat{\sigma}_{ab}^{(s)}$ given in Eq.~(\ref{eq:xsec-soft}) 
receives an additional singular contribution from 
the soft gluon parts of the diagonal splitting functions
$\Delta P_{qq}(x)$ and $\Delta P_{gg}(x)$ in the factorization procedure,
which is proportional to $\delta(1-x)$. 
Only then, all remaining singularities
cancel in the sum of $d\Delta\hat{\sigma}_{ab}^{(s)}$ and $d\Delta\hat{\sigma}_{ab}^{(v)}$,
and the full expression for the subprocess cross section $d\Delta\hat{\sigma}_{ab}$
in Eq.~(\ref{eq:xsec-decomp}) is finite in the limit $\varepsilon\rightarrow 0$.

The numerical evaluation of (\ref{eq:xsec-fact}) for different measurement
functions ${\cal{S}}$ can now be done in parallel with standard Monte Carlo techniques
by randomly generating a large sample of final-state configurations characterized by
$x_1,x_2,x,y,\theta_1,$ and $\theta_2$.
The $\tilde{\rho}$ and $\omega$-distributions
regulating the singularities in $d\Delta\hat{\sigma}^{(f)}$ and the sum
of $d\Delta\hat{\sigma}^{(c\pm)}$ and $d\Delta\hat{\sigma}^{\tilde{c}}$ 
in Eqs.~(\ref{eq:xsec-finite}), (\ref{eq:collinear}), and (\ref{eq:xsec-counter}), respectively,
need special attention. 
To this end, one inserts the definitions of distributions, given in Eq.~(\ref{eq:distrib})
of the Appendix, into Eqs.~(\ref{eq:collinear})-(\ref{eq:xsec-counter}) and computes 
for each phase-space point a set of six correlated weights to account for all possible 
configurations with $x=1$ and $y=\pm1$. 
The values of the measurement functions ${\cal{S}}$ one is interested in are then
multiplied by the appropriate weights and accumulated in different histograms.
In principle, the choice for the parameters $\tilde{\rho}\in [\rho,1[$ and $w\in]0,2[$ 
in (\ref{eq:distrib}) does not matter as it only leads to different values 
for each of the individual, unphysical contributions 
at ${\cal{O}}(\alpha_s^3)$ on the right-hand-side of Eq.~(\ref{eq:xsec-decomp}) but not for their sum.
Large cancellations among the different terms in (\ref{eq:xsec-decomp}) can take place, however,
if $\tilde{\rho}$ is chosen too close to 1 or $\omega$ too close to 0 \cite{ref:mnr}.

%
%
\begin{figure}[th]
\vspace*{-0.35cm}
\includegraphics[width=0.52\textwidth]{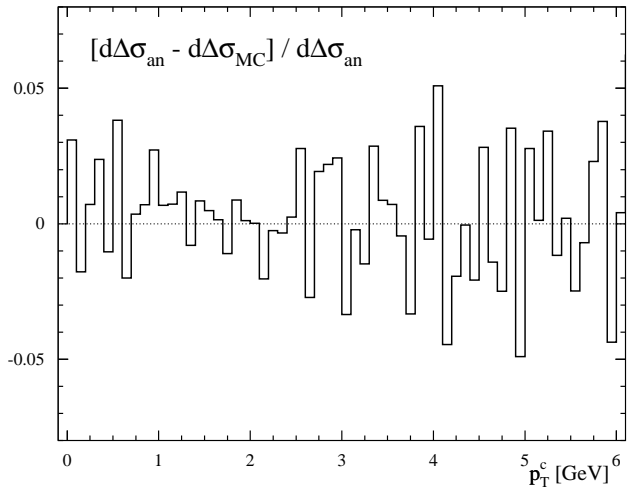}
\vspace*{-0.35cm}
\caption{\label{fig:figure1}
Comparison of the results of our Monte Carlo code, $d\Delta\sigma_{MC}$,
with the analytical calculation $d\Delta\sigma_{\mathrm{an}}$ 
of Refs.~\cite{ref:nlo-pol,ref:hans} for single-inclusive charm production 
as a function of transverse momentum $p_T^c$ and integrated over all rapidities.}
\end{figure}
To validate the numerical implementation of (\ref{eq:xsec-fact}), we compare the 
results obtained with the Monte Carlo techniques outlined above 
to those of the largely analytical code developed and 
used in Refs.~\cite{ref:nlo-pol,ref:hans}. 
Since the analytical calculation is only applicable for single-inclusive heavy quark
(or antiquark) hadroproduction, the comparison is done for charm production 
in longitudinally polarized $pp$ collisions at a c.m.s.\ energy of $\sqrt{S}=200\,\mathrm{GeV}$, 
without any experimental acceptance cuts.
Figure~\ref{fig:figure1} shows the difference of the numerical results obtained with both codes,
labelled as  $d\Delta\sigma_{MC}$ and $d\Delta\sigma_{an}$, as a function of
the transverse momentum $p_T^c$ of the charm quark, normalized to the
analytical calculation $d\Delta\sigma_{an}$.
As can be seen, deviations are at a level of a few per cent at most, well within
the precision of the Monte Carlo integration for the relatively 
small sample of phase-space points used in Fig.~\ref{fig:figure1}.

Needless to say that all discussions in this Section also apply in the unpolarized
case, and the corresponding expressions are obtained by appropriately replacing
all helicity-dependent terms by their spin-averaged counterparts. We fully agree
with the results given in \cite{ref:mnr}.

\section{Phenomenological Studies}
%
\subsection{Preliminaries\label{sec:sec3-a}}
%
Having laid out the technical framework in the previous Section, 
we now turn to a detailed phenomenological study of heavy flavor
hadroproduction in longitudinally polarized $pp$ collisions 
and their semi-leptonic decays at RHIC.
For comparison and to compute experimentally relevant double-spin asymmetries,
defined as
\begin{equation}
\label{eq:all}
A_{LL}\equiv \frac{d\Delta\sigma}{d\sigma}\,\,,
\end{equation}
we also present results for the corresponding unpolarized quantities.
We study the impact of the NLO QCD corrections on the polarized and unpolarized 
cross sections and quantify the theoretical uncertainties from different choices for
unphysical factorization and renormalization scales, 
heavy quark masses, and parameters describing the hadronization of the heavy quarks.

We concentrate on observables of immediate relevance for the RHIC spin program
with collisions of longitudinally polarized protons at a c.m.s.\ energy
of $\sqrt{S}=200\,\mathrm{GeV}$.
These are single-inclusive transverse momentum distributions of electrons and 
muons from semi-leptonic decays of charm and bottom quarks, and, in particular,
invariant mass spectra for two leptons observed in coincidence.
Such measurements have been already carried out in spin-averaged $pp$ collisions
at RHIC \cite{ref:phenix-hq,ref:star-hq} and are intended with longitudinally polarized beams 
once sufficient statistics has been accumulated \cite{ref:phenix-polhq}.

We note that the leptons can stem from both charm and bottom quark decays which cannot
be separated experimentally until displaced vertex detector upgrades have been installed.
Therefore, our results always refer to the sum of charm and bottom production,
their hadronization into $D$ and $B$ mesons, including 
$c\rightarrow D$, $b\rightarrow B$, and ``cascade'' $b\rightarrow B\rightarrow D$
contributions, and the subsequent semi-leptonic decays of the heavy mesons 
into the observed leptons. 
We assume that electrons and muons are detected at central and forward rapidities, 
$|\eta_e|\le 0.35$ and $1.2\le|\eta_{\mu}|\le 2.2$, respectively, which corresponds
to the acceptance of the PHENIX experiment \cite{ref:phenix-hq,ref:phenix-polhq}.

The fragmentation of the heavy quarks into $D$ and
$B$ mesons, i.e., $D^{c\rightarrow D}$ and $D^{b\rightarrow B}$, is modeled by
phenomenological functions extracted from fits to $e^+e^-$ data \cite{ref:frag-review}.
$D$ and $B$ indicate a generic admixture of charm and bottom mesons.
Contrary to fragmentation functions for light quarks and gluons into light mesons \cite{ref:dss},
the non-perturbative functions describing the hadronization of heavy quarks are
very hard, i.e., charm and bottom quarks only lose very little momentum when hadronizing.
The main effect of the fragmentation functions is to introduce a shift in the
normalization of the heavy meson spectra. It depends mainly on the average momentum fraction $z$
taken by the meson, while the details of the shape of $D^{Q\rightarrow H_Q}(z)$ 
have a negligible effect \cite{ref:frag-review}.
One can expect that ratios of cross sections, like in the experimentally most relevant 
double-spin asymmetry (\ref{eq:all}), are much less affected by the actual choice
of $D^{Q\rightarrow H_Q}(z)$.
We use the functional form of Kartvelishvili-Likhoded-Petrov \cite{ref:kart} 
with a single parameter $\alpha_Q$ controlling the hardness of 
\begin{equation}
\label{eq:kart}
D^{Q\rightarrow H_Q}(z)=N_{Q} z^{\alpha_{Q}} (1-z)\,,
\end{equation} 
where $N_{Q}=(\alpha_Q+1)(\alpha_Q+2)$ to normalize the integral of $D^{Q\rightarrow H_Q}(z)$ to one.
We take $\alpha_c=5$ and $\alpha_b=15$ from Tab.~4 in Ref.~\cite{ref:frag-review} as the
default values in Eq.~(\ref{eq:kart}) and vary them in the
range $3\le \alpha_c \le 7$ and $10\le \alpha_b\le 20$, respectively, to estimate the 
uncertainties associated with the choice of $\alpha_Q$.
As in Ref.~\cite{ref:cacciari}, the fragmentation is numerically performed by rescaling the 
heavy quark's three-momentum by $z$ at a constant angle in the laboratory frame,
i.e., $\vec{p}_{H_{Q}}=z\vec{p}_Q$.
The uncertainty introduced by this particular choice for the ``scaling variable'' $z$, which
is not uniquely defined for $D^{Q\rightarrow H_Q}$,
was shown to be not larger than scale and mass uncertainties \cite{ref:cac2} 
and will be not considered further.

The subsequent semi-leptonic decay of the $D$ and $B$ mesons into leptons is controlled
by another set of phenomenological functions $f^{H_Q\rightarrow e,\mu}$ which need
to be extracted from data as well. Here we use the spectra obtained in
Ref.~\cite{ref:cacciari,ref:vogt} based on BaBar and CLEO data \cite{ref:babar-cleo}. 
We note that we do {\em not} normalize 
our cross sections with the appropriate branching ratios for 
$D\rightarrow e$, $B\rightarrow e$, etc.,
which are all close to 10$\%$ \cite{ref:pdg}. Of course, branching ratios
drop out of experimentally relevant double-spin asymmetries (\ref{eq:all}).

The main motivation to study heavy flavor production with 
polarized beams at RHIC is the expected sensitivity to the 
helicity-dependent gluon density through the tree-level gluon-gluon fusion process,
$gg\rightarrow Q\bar{Q}$, which is known to be dominant for unpolarized collisions
up to the largest values of the heavy quark's transverse momentum
currently accessible at RHIC \cite{ref:nlo-unpol,ref:nlo-unpol2}.

We will show, however, that the fractional contribution of 
gluon-gluon fusion to the spin-dependent cross section 
depends crucially on the assumed set of polarized parton densities.
Our default choice is the DSSV set \cite{ref:dssv}, obtained in a global QCD analysis
of the latest spin-dependent data, including those from RHIC on single-inclusive
pion and jet production \cite{ref:phenix,ref:star}. 
Due to the smallness of $\Delta g(x)$ in the DSSV set
and a node in the $x$-shape near $x\simeq 0.1$ \cite{ref:dssv}, 
the $q\bar{q}$ annihilation subprocess, $q\bar{q}\rightarrow Q\bar{Q}$, turns 
out to be the dominant mechanism for 
charm and bottom production in polarized $pp$ collisions at RHIC.
This is in sharp contrast to naive
expectations based on unpolarized results.
For comparison and to study the sensitivity to $\Delta g(x)$,
we adopt also two alternative sets of spin-dependent parton densities, 
GRSV(std)~\cite{ref:grsv} and DNS(KRE)~\cite{ref:dns}, 
both characterized by a positive gluon polarization of moderate size.
In general, for gluon polarizations from current QCD fits
\cite{ref:dssv,ref:grsv,ref:dns} , the double-spin asymmetries
for leptons from heavy flavor decays all turn out to be small, often
well below the one percent level, making their measurement very challenging. 
This is in particular true for single-inclusive lepton observables; see below.

Heavy flavor production at RHIC cannot compete with the 
statistical precision achievable for more abundant probes 
of the nucleon's spin structure, like pions and jets \cite{ref:phenix,ref:star},
which are already used in global fits \cite{ref:dssv}. 
Nevertheless, measurements of double-spin asymmetries related to heavy flavor production
will be crucial for further testing and establishing the all important
concept of factorization and universality for helicity-dependent
scattering processes and parton densities, respectively.
The underlying dynamics of the partonic scattering processes, i.e., $gg\rightarrow Q\bar{Q}$
and $q\bar{q}\rightarrow Q\bar{Q}$, is very much different as compared
to the multitude of QCD processes driving the 
production of light hadrons \cite{ref:nlo-pions} or jets \cite{ref:nlo-jets}.

In the computation of the NLO unpolarized cross sections in (\ref{eq:all}),
which proceeds along similar lines as outlined in Sec.~II, for details,
see Ref.~\cite{ref:mnr}, we use the NLO CTEQ6M parton densities \cite{ref:cteq6} 
and values for the strong coupling $\alpha_s$.
Since the DSSV analysis \cite{ref:dssv} does not provide a 
LO set of spin-dependent parton distributions,
our LO results always refer to the Born part of the full
NLO calculation, i.e., they are computed with NLO parton densities and values
for $\alpha_s$. Strictly speaking this is, of course, inconsistent as it introduces some
unwanted scheme dependence into a tree-level quantity. Nevertheless, the LO results
should give a faithful estimate of the relevance of NLO corrections. 
As will be demonstrated below, they turn out to be sizable and rather different
for unpolarized and polarized cross sections such that they do not cancel in
experimentally relevant double-spin asymmetries.

We take $m_{c}=1.35\,\mathrm{GeV}$ and $m_{b}=4.75\,\mathrm{GeV}$ as reference
values for the charm and bottom quark mass and vary them
in the range $1.2\le m_{c} \le 1.5\,\mathrm{GeV}$ and
$4.5\le m_{b} \le5.0\,\mathrm{GeV}$, respectively, to estimate the resulting mass uncertainties.
For the factorization and renormalization scale
we take $\mu_f=\mu_r=\xi(m_Q^2+[(p_T^Q)^2+(p_T^{\bar{Q}})^2]/2)^{1/2}$ 
with $\xi=1$ as the central value. 
As usual, the sensitivity of the cross section to missing higher order
corrections is estimated by varying $\mu_f$ and $\mu_r$ simultaneously 
in the range $1/2\le \xi\le 2$.
Following the procedure 
used for unpolarized charm and bottom production at RHIC in Ref.~\cite{ref:cacciari} , 
we also vary $\mu_f$ and $\mu_r$ independently in the same range of $\xi$ and combine
the ensuing uncertainty with the one stemming from variations of $m_{c,b}$ in quadrature.
Unless stated otherwise, we use the central values for $\mu_f$, $\mu_r$,
$m_{c,b}$, and $\alpha_{c,b}$ given above.

\subsection{Heavy Flavor Cross Sections And Correlations\label{sec:scale}}
%
%
\begin{figure}[th]
\vspace*{-0.35cm}
\includegraphics[width=0.52\textwidth]{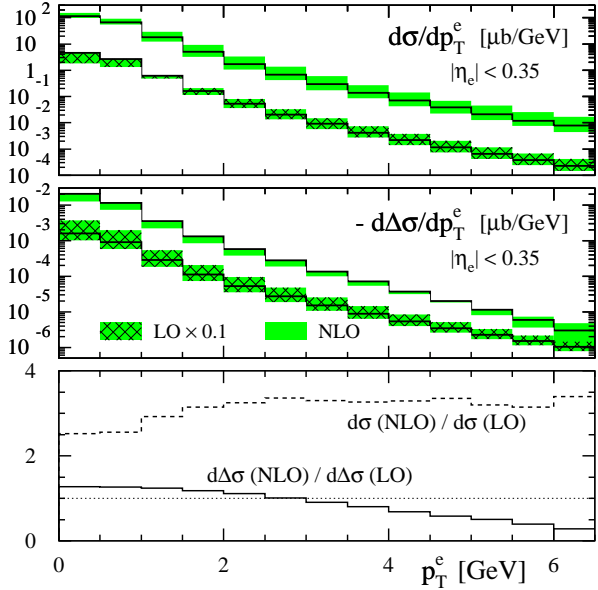}
\vspace*{-0.35cm}
\caption{\label{fig:figure2}
Scale dependence of the single-inclusive transverse momentum spectrum of 
electrons from heavy quark decays at central rapidities $|\eta_e|\le0.35$
in unpolarized (upper panel) and polarized (middle panel)
$pp$ collisions at RHIC.
All scales are varied simultaneously, i.e.,  
$\mu_f=\mu_r=\xi(m_Q^2+[(p_T^Q)^2+(p_T^{\bar{Q}})^2]/2)^{1/2}$,
in the range $1/2\le\xi\le2$ (shaded bands). The solid lines correspond to the default
choice $\xi=1$.
Note that the LO results are rescaled by a factor of 0.1, and all polarized 
cross sections are multiplied by $-1$.
The lower panel shows the ratio of NLO to LO polarized
and unpolarized cross sections ($K$-factor).}
\end{figure}

We begin our detailed numerical studies with a discussion of 
unpolarized and polarized cross sections for various
decay lepton distributions accessible at RHIC.

Figure~\ref{fig:figure2} shows the single-inclusive transverse momentum
spectrum of electrons from charm and bottom decays in LO and NLO
accuracy, integrated over the angular acceptance of the PHENIX 
detector, i.e., $|\eta_e|\le 0.35$ \cite{ref:phenix-hq}. 
Similar results are obtained for the STAR experiment \cite{ref:star-hq} with its larger 
acceptance for electrons at central rapidities, $|\eta_e|<1$,
and hence not shown.
The transverse momentum $p_T^e$ is limited to a region which should be accessible
with luminosities envisaged in longitudinally polarized $pp$ collisions at RHIC.
Photon conversion, $\gamma\rightarrow e^+e^-$, and $\pi^0\rightarrow \gamma e^+e^-$
Dalitz decays are the dominant source of electron background for such measurements
and may require an additional cut $p_T^e>1\,\mathrm{GeV}$ \cite{ref:phenix-hq,ref:star-hq}.
Recall that the branching ratios of about $10\%$ are not included in
the cross sections shown in Fig.~\ref{fig:figure2}.

The solid lines are obtained with the default values of the heavy
quark masses, scales, parameters, and parton densities stated in the
previous Subsection.
The shaded bands indicate the theoretical uncertainty from varying the 
factorization and renormalization scale simultaneously in the range $1/2\le \xi\le 2$
specified above.
Note that the LO results are rescaled by a factor of 0.1, and all polarized 
cross sections are multiplied by -1 to display them on a logarithmic scale.
The bottom panel of Fig.~\ref{fig:figure2} gives the resulting unpolarized
and polarized ``$K$-factors'', defined as usual by the ratio
\begin{equation}
\label{eq:kfac}
K\equiv\frac{d(\Delta)\sigma^{\mathrm{NLO}}}{d(\Delta)\sigma^{\mathrm{LO}}}\,\,.
\end{equation}
One notices that the NLO corrections are sizable in the unpolarized case,
$K\simeq 3$, but moderate for polarized $pp$ collisions, except for the region
$p_T^e\gtrsim 5\,\mathrm{GeV}$. Here, the polarized cross section approaches a node,
and perturbative corrections are artificially enhanced.

%
%
\begin{figure}[th]
\vspace*{-0.35cm}
\includegraphics[width=0.52\textwidth]{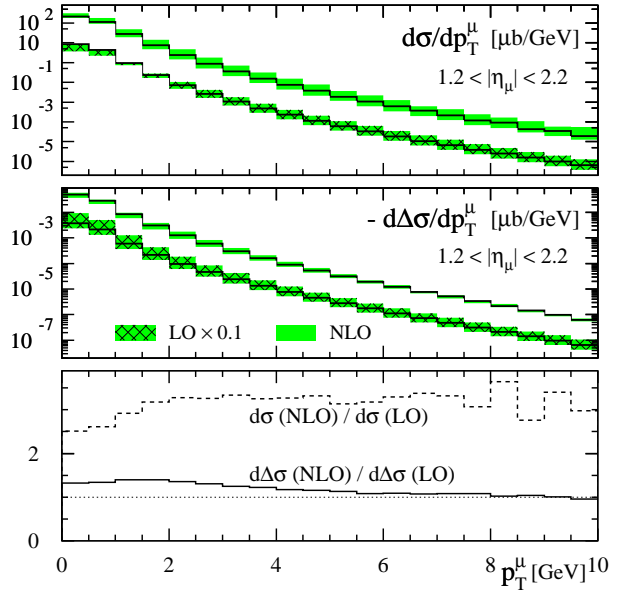}
\vspace*{-0.35cm}
\caption{\label{fig:figure3}
Same as in Fig.~\ref{fig:figure2} but for the 
single-inclusive transverse momentum spectrum of 
muons from heavy quark decays at forward rapidities 
$1.2\le|\eta_{\mu}|\le2.2$.}
\end{figure}
Less pronounced NLO corrections for polarized cross sections are a 
rather generic feature and have been observed already for other hadronic processes
such as single-inclusive pion \cite{ref:nlo-pions} and jet \cite{ref:nlo-jets} production. 
To some extent this behavior can be traced back to the less singular scale
evolution of polarized parton densities at small momentum fractions $x$ \cite{ref:nlo-split}.
This has the effect 
that the partonic threshold region, which is the source of large logarithmic corrections 
associated with the emission of soft gluons, is less emphasized
in the convolution (\ref{eq:xsec-fact}) than in the unpolarized case.
Specifically for heavy flavor production, it was noticed in \cite{ref:nlo-pol} that 
large NLO corrections to the gluon-gluon fusion process
related to amplitudes with a gluon exchange in the $t$-channel
are independent of the helicities of the interacting gluons and hence do not contribute
to the polarized cross section.
Substantially different $K$-factors for unpolarized and polarized cross sections
immediately imply that Born level estimates for double-spin asymmetries (\ref{eq:all})
can serve only as very rough estimates. In general, they are insufficient for any quantitative
analysis such as a global QCD extraction of spin-dependent parton densities.

%
%
\begin{figure}[th]
\vspace*{-0.35cm}
\includegraphics[width=0.52\textwidth]{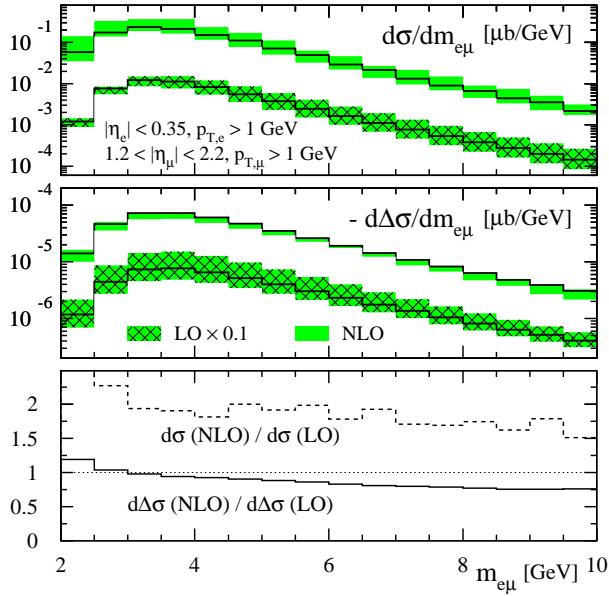}
\vspace*{-0.35cm}
\caption{\label{fig:figure4}
Same as in Fig.~\ref{fig:figure2} but for the 
invariant mass spectrum of electron-muon correlations from heavy quark decays.
Electrons are detected at central rapidities, $|\eta_e|\le0.35$, and
muons are detected at forward rapidities, $1.2\le|\eta_{\mu}|\le2.2$.
The transverse momenta of both electrons and muons are required to be
larger than $1\,\mathrm{GeV}$.}
\end{figure}
As can be inferred from Fig.~\ref{fig:figure2}, theoretical ambiguities due
to the choice of $\mu_r$ and $\mu_f$ in (\ref{eq:xsec-fact}) 
are reduced in the polarized but not in the unpolarized case.
Along with the observation of large QCD corrections, this indicates the
need for next-to-next-to-leading order corrections for the unpolarized
cross section to better control the dependence on $\mu_f$ and $\mu_r$.
We note that substantial progress toward this direction 
has already been made recently \cite{ref:hq-nnlo},
mainly to allow for precision studies with the large amount of
top quarks expected to be produced at the CERN-LHC.
In addition, fixed order calculations need to be amended by all-order
resummations if $\ln p_T^Q/m_Q$ becomes large. This was achieved in \cite{ref:resum} 
but is not really relevant for our discussions here 
since we are mainly interested in the region where $p_T^Q\sim m_Q$.
We postpone a discussion of theoretical uncertainties due to the choice of
$m_Q$ and $\alpha_Q$ in Eq.~(\ref{eq:kart}), as well as the effect of 
varying $\mu_r$ and $\mu_f$ independently, until the end of this Subsection.

The single-inclusive transverse momentum spectrum of muons from
heavy quark decays is shown in Fig.~\ref{fig:figure3} in 
LO and NLO accuracy.
The pseudorapidity $\eta_{\mu}$ of the muon is integrated 
in the range $1.2\le|\eta_{\mu}|\le2.2$ corresponding to the
angular acceptance of the PHENIX experiment.
All observations made in Fig.~\ref{fig:figure2} 
regarding the relevance of NLO corrections, the behavior of
the $K$-factor, and the dependence on $\mu_{f,r}$ apply also here.
The polarized $K$-factor stays even closer to one 
than in Fig.~\ref{fig:figure2} as
$d\Delta\sigma/dp_T^{\mu}$ develops no node in the
$p_T^{\mu}$ range shown.

%
%
\begin{figure}[th]
\vspace*{-0.35cm}
\includegraphics[width=0.52\textwidth]{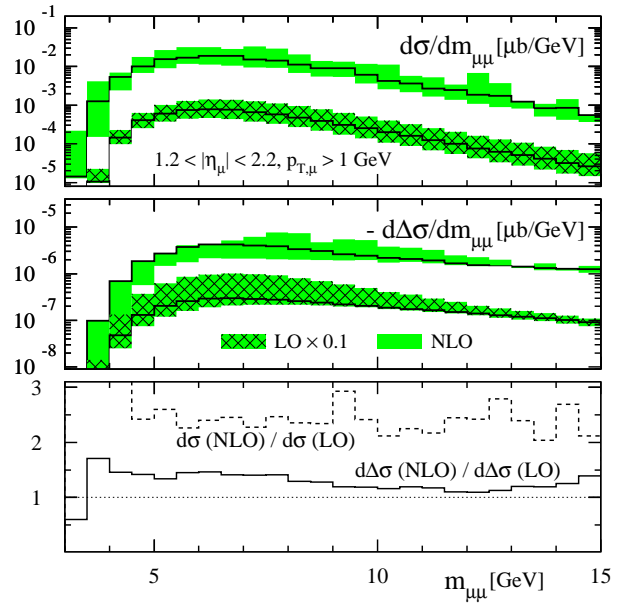}
\vspace*{-0.35cm}
\caption{\label{fig:figure5}
Same as in Fig.~\ref{fig:figure2} but for the 
invariant mass spectrum of muon-muon correlations from heavy quark decays.
Both muons are detected at forward rapidities, $1.2\le|\eta_{\mu}|\le2.2$,
but in different hemispheres. The transverse momentum of each muons 
is required to be larger than $1\,\mathrm{GeV}$.}
\end{figure}
By glancing at the relative sizes of the unpolarized and polarized 
single-inclusive transverse momentum spectra for electrons and muons shown
in Fig.~\ref{fig:figure2} and \ref{fig:figure3}, respectively, 
it becomes immediately obvious that the corresponding double-spin asymmetries
$A_{LL}^e$ and $A_{LL}^{\mu}$, to which we turn to in Subsec.~\ref{sec:all}, 
are very small if the most up-to-date DSSV parton densities \cite{ref:dssv}
are used.
Expected asymmetries of the order of a few tenths
of a percent are extremely challenging experimentally as
systematic uncertainties, like from the determination of
the relative beam luminosities at RHIC, are of similar size \cite{ref:spinplan}.
At higher $p_T^{e,\mu}$, where double-spin asymmetries are largest,
the single-inclusive cross sections in Figs.~\ref{fig:figure2} and \ref{fig:figure3}
have dropped already several
orders of magnitude from their peak values, 
and measurements require substantial integrated luminosities.

More promising appear to be observables where
both the heavy quark and the heavy antiquark decay semi-leptonically,
and both leptons are observed in coincidence.
This is also where our numerical phase-space integration
and the flexible Monte Carlo code for polarized heavy flavor hadroproduction
introduced in Sec.~\ref{sec:tech} become truly essential.
Particle correlations are hard, and often impossible, to compute at NLO
with largely analytical methods, see, e.g., \cite{ref:christof}.

Figure~\ref{fig:figure4} shows our results for the invariant mass spectrum
of  electron-muon correlations from semi-leptonic decays of $D$ and $B$ mesons
within the angular acceptance of the PHENIX detector, i.e.,
$|\eta_e|\le0.35$ and $1.2\le|\eta_{\mu}|\le2.2$. In addition, we require
a minimum transverse momentum for both leptons of $1\,\mathrm{GeV}$ as required
by experiment.
As in Figs.~\ref{fig:figure2} and \ref{fig:figure3},
results shown as solid lines are obtained with the default 
choice of parameters. Again, shaded bands indicate the theoretical uncertainty 
from varying the factorization and renormalization scales simultaneously 
in the range $1/2\le \xi\le 2$.

%
%
\begin{figure}[th]
\vspace*{-0.35cm}
\includegraphics[width=0.52\textwidth]{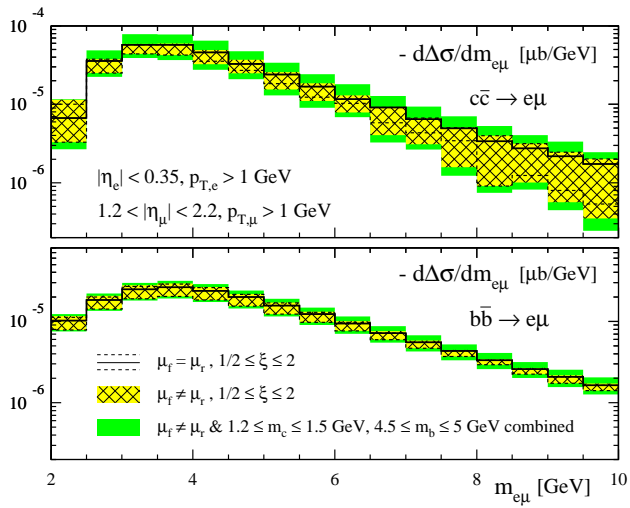}
\vspace*{-0.35cm}
\caption{\label{fig:figure5a}
Scale and mass uncertainties for the polarized invariant mass spectrum
of electron-muon correlations at NLO accuracy, using the same cuts
as in Fig.~\ref{fig:figure4}. The upper and lower panel shows the contribution
from $c\bar{c}$ and $b\bar{b}$ decays, respectively, multiplied by $-1$.
The dashed lines indicate the range of uncertainties for 
$\mu_f=\mu_r$, $1/2\le \xi \le 2$, and central values of $m_{c,b}$ as in Fig.~\ref{fig:figure4}.
The solid curves are for $\xi=1$. 
The effect of varying $\mu_f$ and $\mu_r$ independently in the same range of $\xi$ 
for fixed $m_c=1.35\,\mathrm{GeV}$, $m_b=4.75\,\mathrm{GeV}$ 
and for $1.35\le m_c\le 1.5\,\mathrm{GeV}$, $4.5\le m_b\le 5.0\,\mathrm{GeV}$
is illustrated by hatched and solid bands, respectively. In the latter case,
scale and mass uncertainties are combined in quadrature (see text).} 
\end{figure}
The $K$-factor, shown in the lower panel of Fig.~\ref{fig:figure4}, is smaller
than what was found for the single-inclusive observables 
in Figs.~\ref{fig:figure2} and \ref{fig:figure3} in the unpolarized case.
Still, NLO corrections differ considerably for the unpolarized and polarized 
invariant mass spectra. Again, the corrections are such that the 
corresponding double-spin asymmetry is reduced at NLO accuracy.
The scale uncertainty is significantly smaller for the spin-dependent cross section
with NLO corrections included. The improvement in the helicity-averaged case
is much less pronounced.

Compared to the single-inclusive results in Figs.~\ref{fig:figure2} and \ref{fig:figure3}, 
the cross sections obtained for the electron-muon invariant mass spectrum are smaller,
but $d(\Delta)\sigma/dm_{e\mu}$ drops much less with increasing $m_{e\mu}$ than 
$d(\Delta)\sigma/dp_T^{e,\mu}$ with increasing $p_T^e$ or $p_T^{\mu}$. 
This makes measurements of $A_{LL}^{e\mu}$ at comparatively large values of $m_{e\mu}$ feasible.
We do not consider here correlations with back-to-back electrons at central
rapidities. Electron-muon correlations are phenomenologically more
interesting due to their asymmetric kinematics with respect to rapidity, 
probing the interacting partons at different momentum fractions $x$ as will be demonstrated below.

The corresponding invariant mass spectrum for two muons from $D$ and $B$ meson
decays observed in coincidence is
shown in Fig.~\ref{fig:figure5}. Both muons are required to
have $1.2\le|\eta_{\mu}|\le2.2$ and $p_T^{\mu}>1\,\mathrm{GeV}$, with one muon
detected at forward (positive) and one muon detected at backward (negative)
pseudorapidities.
Again, the cross sections $d(\Delta)\sigma$ decrease rather slowly with increasing invariant
mass $m_{\mu\mu}$. This observable is very demanding in terms of required
Monte Carlo statistics as can be seen by the still fairly pronounced fluctuations, 
most noticeable in the unpolarized $K$-factor.
The general trend and features of the cross sections are, however, reliable.
As before, NLO corrections are more significant in the unpolarized case, where
$K\simeq 2$. Unfortunately, the reduction of the theoretical ambiguities related to the
choice of $\mu_{f,r}$ is only marginal at NLO.

We now turn to a more detailed discussion of theoretical uncertainties for the
observables discussed in this Section, taking the phenomenologically interesting
invariant mass spectrum for electron-muon correlations, presented in Fig.~\ref{fig:figure4}, 
as an example.
Qualitatively very similar results are obtained for the other cross sections given
in Figs.~\ref{fig:figure2}, \ref{fig:figure3}, and  \ref{fig:figure5} and hence
not shown here.

The impact of varying $\mu_f$ and $\mu_r$ independently is shown in Fig.~\ref{fig:figure5a}
for the electron-muon invariant mass spectrum in polarized $pp$ collisions at NLO accuracy.
Since we are also interested in variations of $m_{c,b}$, 
the contribution from $c\bar{c}$ and $b\bar{b}$ decays are shown in separate panels and
add up to $-d\Delta\sigma/dm_{e\mu}$ discussed in Fig.~\ref{fig:figure4}.
Following Ref.~\cite{ref:cacciari}, we compute our results for seven different
settings of scales $\mu_{f,r}=\xi_{f,r}(m_Q^2+[(p_T^Q)^2+(p_T^{\bar{Q}})^2]/2)^{1/2}$,
using $(\xi_f,\xi_r) = \{(1,1),(2,2),(1/2,1/2),(1,1/2),(2,1),(1/2,1),(1,2)\}$
and keeping $m_{Q}$ fixed to their central values $m_c=1.35\,\mathrm{GeV}$ and
$m_b=4.75\,\mathrm{GeV}$.
The envelope of all resulting curves defines the scale uncertainty and 
is show as hatched bands in Fig.~\ref{fig:figure5a}.
For comparison, the dashed lines indicate the range of uncertainties for
standard choice $\mu_f=\mu_r$ used in Fig.~\ref{fig:figure4}.
As can be seen, taking $\mu_f\neq\mu_r$ does not significantly enlarge
the scale ambiguities for the polarized cross section, in particular, 
for the contribution from bottom quarks shown in the lower panel.

%
%
\begin{figure}[th]
\vspace*{-0.35cm}
\includegraphics[width=0.52\textwidth]{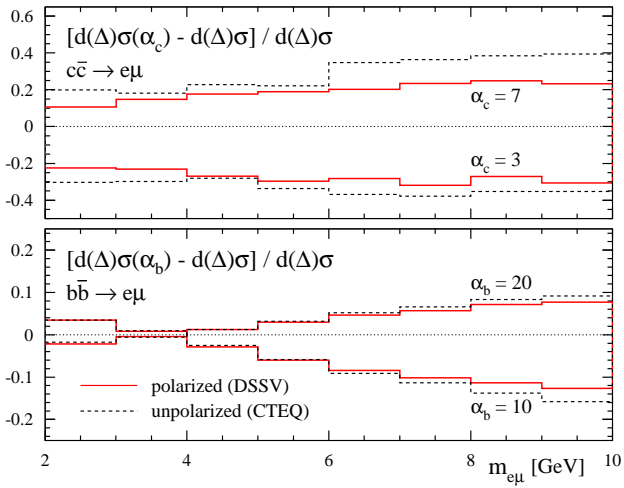}
\vspace*{-0.35cm}
\caption{\label{fig:figure5b}
Dependence of the polarized (solid lines) and unpolarized (dashed lines)
invariant mass spectra for electron-muon correlations at NLO accuracy
on the choice of fragmentation parameters $\alpha_c$ (upper panel)
and $\alpha_b$ (lower panel) defined in Eq.~(\ref{eq:kart}).
Displayed are the relative deviations for $c\bar{c}$ and $b\bar{b}$ decays
using $\alpha_c=3,7$ and $\alpha_b=10,20$
with respect to the cross sections obtained for our 
default values $\alpha_c=5$ and $\alpha_b=15$ \cite{ref:frag-review},
respectively. $\mu_f$, $\mu_r$, and heavy quark masses are taken at
their central values, and the same experimental cuts as in  Fig.~\ref{fig:figure4}
are adopted.}
\end{figure}
The solid bands in Fig.~\ref{fig:figure5a} take also variations of 
$m_c$ and $m_b$ into account. The recipe we follow here is again similar
to the one used to estimate theoretical uncertainties for unpolarized 
heavy flavor production \cite{ref:cacciari,ref:cac2}.
In practice, we add scale and mass uncertainties in quadrature, and the
envelope of all results is defined by
$C+[(M_{\mu}^+-C)^2+(M_{m_Q}^+-C)^2]^{1/2}$ and
$C-[(C-M_{\mu}^-)^2+(C-M_{m_Q}^-)^2]^{1/2}$.
Here, $C$ denotes the results obtained for central values of scales and masses. 
$M_{\mu}^+$ ($M_{\mu}^-$) are the maximum (minimum) cross sections 
computed for $\mu_f\neq\mu_r$, $m_c=1.35\,\mathrm{GeV}$, and
$m_b=4.75\,\mathrm{GeV}$, as depicted by the hatched bands.
Correspondingly, $M_{m_Q}^+$ ($M_{m_Q}^-$) denote the
maximum (minimum) cross sections for $\xi_{f,r}=1$ and
varying $m_c$ and $m_b$ in the range
$1.35\le m_c\le 1.5\,\mathrm{GeV}$ and
$4.5\le m_b\le 5.0\,\mathrm{GeV}$, respectively.
In general, the combined uncertainties are much smaller for 
$b\bar{b}$ than for $c\bar{c}$ production and decays, which is not
too surprising. In both cases, 
variations of $m_Q$ add noticeably to the theoretical uncertainties.

The dependence of the cross sections
on the choice of $\alpha_Q$ in the non-perturbative function $D^{Q\rightarrow H_Q}(z)$
describing the hadronization of the heavy quarks into $D$ and $B$ mesons, 
see Eq.~(\ref{eq:kart}),
is illustrated in Fig.~\ref{fig:figure5b}. Again, we take the invariant mass spectrum for
electron-muon correlations as an representative example.
We vary $\alpha_c$ and $\alpha_b$ in the range \cite{ref:frag-review} $3\le\alpha_c\le7$ and
$10\le\alpha_b\le 20$, respectively, and show the impact on the invariant mass spectrum
as relative uncertainty with respect to the results obtained for the 
central values $\alpha_c=5$ and $\alpha_b=15$ used in Fig.~\ref{fig:figure4}.

It turns out that polarized and unpolarized invariant mass spectra are 
affected very much in the same way by variations of $\alpha_{c,b}$.
For charm production and taking $3\le\alpha_c\le7$, it roughly amounts to a
shift in the normalization of the cross sections by $\pm\,20\div30\%$.
The impact of varying $\alpha_b$ on the contribution to $d(\Delta)\sigma/dm_{e\mu}$
from bottom decays is significantly smaller, up to about $\pm\,10\%$ 
deviation from the results for $\alpha_b=15$, but is less uniform with $m_{e\mu}$.
These observations have the important implication that theoretical uncertainties associated
with the actual choice of $\alpha_{c,b}$ drop out to a large extent for experimentally 
relevant double-spin asymmetries $A_{LL}$ discussed in Sec.~\ref{sec:all} below.

\subsection{Subprocess, Charm, and Bottom Fractions}
%
%
%
\begin{figure}[th]
\vspace*{-0.35cm}
\includegraphics[width=0.52\textwidth]{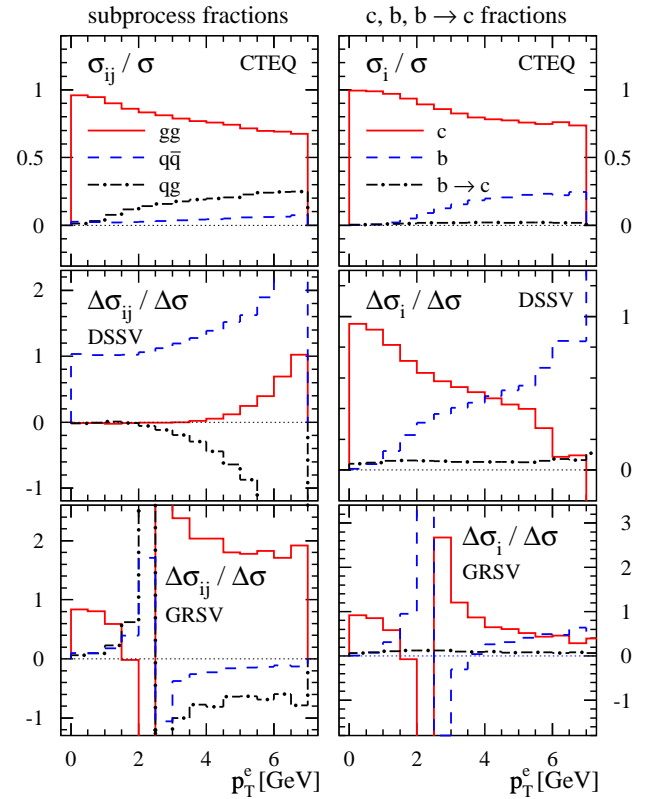}
\vspace*{-0.35cm}
\caption{\label{fig:figure6}
Fractional amount of different partonic subprocesses at NLO accuracy 
(left column) and
of charm, bottom, and cascade ($b\rightarrow c$) decays (right column)
contributing to the single-inclusive transverse momentum spectrum of electrons
shown in Fig.~\ref{fig:figure2}. 
Results are shown for unpolarized (upper row) and polarized (middle and lower rows)
$pp$ collisions at RHIC using the CTEQ6 \cite{ref:cteq6}, DSSV \cite{ref:dssv}, 
and GRSV \cite{ref:grsv} set of parton densities, respectively.}
\end{figure}
We now take a detailed look at the fractional contributions of the
different partonic hard scattering processes to the cross sections shown
in Figs.~\ref{fig:figure2} - \ref{fig:figure5}. This will help to
understand the dependence of the double-spin asymmetries on different
sets of polarized parton densities, to be discussed in the next Subsection.
Since charm and bottom decays both contribute to the lepton spectra,
we also present their fractional contributions. This includes
also the ``cascade'' decay $b\rightarrow c \rightarrow e,\mu$, which
is modeled following the procedure discussed in \cite{ref:cacciari}. It
is found to be negligible for all observables we are interested in.

The left-hand-side of Fig.~\ref{fig:figure6} shows the contributions 
of the three different subprocesses at NLO accuracy,
with $gg$, $q\bar{q}$, and $qg$ initial-states, 
to the single-inclusive decay electron spectra shown in Fig.~\ref{fig:figure2}. 
In the unpolarized case (upper panel), gluon-gluon fusion is the by far
dominant subprocess for heavy flavor production at RHIC energies, with
$q\bar{q}$ annihilation becoming somewhat more relevant at larger values of 
transverse momentum $p_T^e$. 
Interestingly enough, the genuine NLO, i.e., $\alpha_s$ suppressed, 
$qg$ scattering process also contributes very significantly at larger $p_T^e$,
even exceeding the $q\bar{q}$ annihilation cross section.
This observation can be linked to the abundance of gluons at
all momentum fractions $x$ \cite{ref:cteq6}. This implies that the partonic flux 
relevant for $qg$ scattering, i.e., $q(x_1,\mu_f)g(x_2,\mu_f)$, is much
larger than the corresponding flux for $q\bar{q}$ annihilation, in particular,
at the medium-to-large momentum fractions $x_{1,2}$ relevant for RHIC. 
This compensates for the ${\cal{O}}(\alpha_s)$ suppression 
in the $qg$ hard scattering channel. 
In $p\bar{p}$ scattering, e.g., at the TeVatron, 
where antiquarks are ``valence'' quarks in the antiproton beam,
this is different, and the $q\bar{q}$ flux is much enhanced.
A similar observation concerning the relevance of the 
$q\bar{q}$ annihilation channel was made also 
for fixed-target experiments in Ref.~\cite{ref:hans},
where it is expected to contribute very significantly 
to charm hadroproduction in proposed  $p\bar{p}$ collisions at the GSI-FAIR facility, 
but not in $pp$ scattering of similar c.m.s.\ energy planned at J-PARC.

In general, the situation is much more involved in the polarized case, 
where both hard scattering cross sections and parton densities
are not positive definite and can contribute with either sign, depending
on the kinematics relevant for a particular process.
In the vicinity of sign changes, large cancellations are to be expected.
As we shall demonstrate below, depending on the chosen set of 
polarized parton densities, the subprocess fractions can differ 
considerably from each other and often
gluon-gluon fusion does not dominate, in contrast to the unpolarized case.

The middle panel of Fig.~\ref{fig:figure6} shows our results
for the polarized subprocess fractions obtained with the DSSV set
\cite{ref:dssv}, our default choice of parton densities 
used in Figs.~\ref{fig:figure2}-\ref{fig:figure5}.
At small $p_T^e$, the cross section is
entirely dominated by $q\bar{q}$ annihilation, contrary to the
unpolarized case. 
Towards larger $p_T^e$, both $gg$ and $qg$ processes contribute
significantly but with opposite sign, leading to  
strong cancellations.
This happens, however, in a kinematic region close to a sign change
of the cross section at $p_T^e\simeq 7\,\mathrm{GeV}$.

A rather different pattern of fractional subprocess contributions 
can be found in the bottom panel of 
Fig.~\ref{fig:figure6}, where the GRSV(std) \cite{ref:grsv} parton densities
were used. Note that the cross section has a sign change near
$p_T^e=2\,\mathrm{GeV}$. This explains the complicated behavior of the ratios
in this region and makes it very awkward to display them properly.
Like in the unpolarized case, gluon-gluon fusion
is the most important contribution to the cross section.
At larger $p_T^e$, $q\bar{q}$ and $qg$ subprocesses become more relevant,
both contributing with the opposite sign than $gg$ scattering,
leading again to fairly significant cancellations.

%
%
\begin{figure}[th]
\vspace*{-0.35cm}
\includegraphics[width=0.52\textwidth]{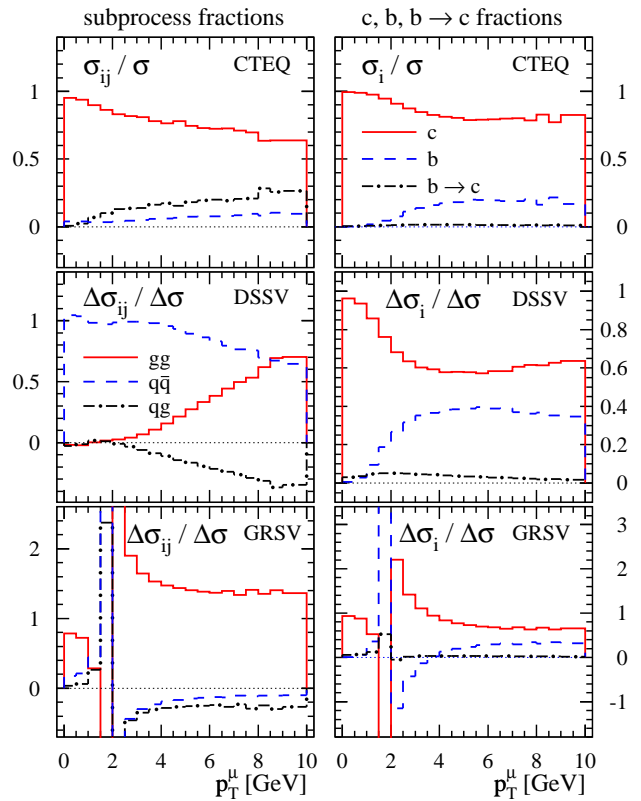}
\vspace*{-0.35cm}
\caption{\label{fig:figure7}
Same as in Fig.~\ref{fig:figure6} but now for the
single-inclusive transverse momentum spectrum of muons
shown in Fig.~\ref{fig:figure3}.}
\end{figure}
The gross features of the results in Fig.~\ref{fig:figure6} obtained 
with DSSV and GRSV parton densities 
can be readily understood by comparing the size and sign of the 
individual quark, antiquark, and gluon densities in both sets, 
see, e.g., Fig.~2 in Ref.~\cite{ref:dssv}.
Since the decay electrons stem from heavy (anti)quarks produced at central
rapidities, the interacting partons have very similar momentum
fractions, i.e., $x_1\simeq x_2$.
Therefore, $\Delta g(x_1)\Delta g(x_2)>0$, irrespective of the node in the
DSSV gluon distribution, and the sign of the $gg$
contribution follows the sign of the hard scattering cross section,
which changes from positive at small $p_T$ to negative at larger $p_T$ values.

%
%
%
\begin{figure}[th]
\vspace*{-0.35cm}
\includegraphics[width=0.52\textwidth]{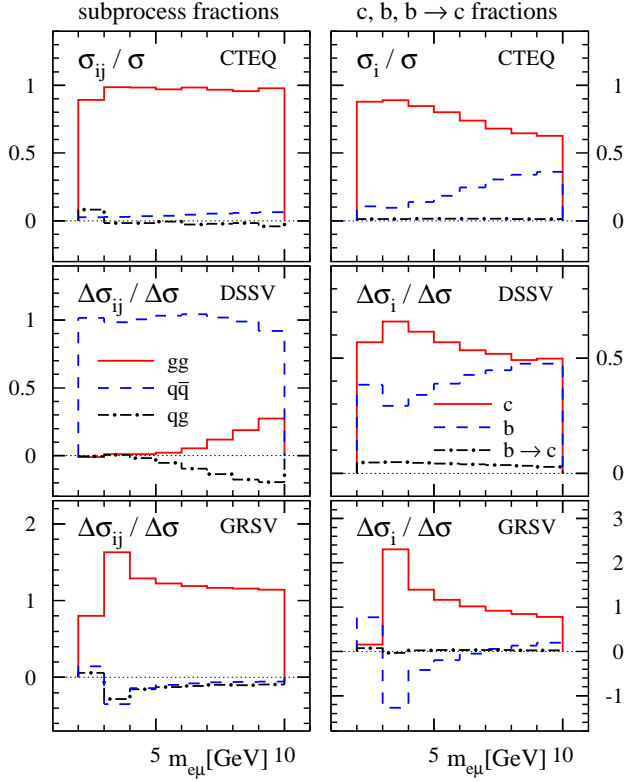}
\vspace*{-0.35cm}
\caption{\label{fig:figure8}
Same as in Fig.~\ref{fig:figure6} but now for the
invariant mass spectrum of electron-muon correlations
shown in Fig.~\ref{fig:figure4}.}
\end{figure}
Due to helicity conservation, 
$d\Delta\hat{\sigma}_{q\bar{q}}=-d\hat{\sigma}_{q\bar{q}}<0$ \cite{ref:nlo-pol}, 
and the sign of the $q\bar{q}$ contribution depends on the individual 
parton densities for each quark and antiquark flavor.
All sets of polarized parton densities have $\Delta u(x)>0$ and $\Delta d(x)<0$,
resembling the features of the naive quark model.
The GRSV(std) set \cite{ref:grsv} assumes an $SU(3)$ symmetric sea, 
with all antiquark polarizations being negative, such
that for the dominant $u$-quarks one has $\Delta u(x_1)\Delta\bar{u}(x_2)<0$, 
resulting in a net positive contribution to the cross section.
This is exactly opposite in the DSSV set \cite{ref:dssv}, 
where $\Delta u(x_1)\Delta\bar{u}(x_2)>0$, unless $x_2$ gets very large. 
The genuine NLO $qg$ subprocess cross section, as well as the sum of all quark and antiquark
polarizations, $\Delta\Sigma=\sum_q [\Delta q+\Delta\bar{q}]$,
are both positive. This implies that the sign of the $qg$ contribution 
depends on the sign of $\Delta g(x)$ in the relevant region of $x$,
which turns out to be positive for both GRSV(std) and DSSV.
Overall, the fractional contributions of the individual subprocesses to the 
single-inclusive decay electron spectrum are essentially controlled by
the modulus of the polarized gluon density, $|\Delta g(x)|$, which is
much larger for the GRSV(std) set, i.e.,  $|\Delta g(x)|_{GRSV}\gg|\Delta g(x)|_{DSSV}$.
The bigger the gluon density, the closer the result is to what we have found
in the unpolarized case. It turns out that even for the moderate gluon polarization
of the GRSV(std) set, the gluon-gluon channel prevails for all $p_T^e$ values shown
in Fig.~\ref{fig:figure6}.

The right-hand-side of Fig.~\ref{fig:figure6} shows the 
fractional contributions of the charm, bottom, and 
``cascade'' $b\rightarrow c$ decays to the single-inclusive 
transverse momentum spectrum of electrons.
For $p_T^e\lesssim 2\,\mathrm{GeV}$, almost all electrons originate
from charm decays, but above the bottom contribution catches
up, yielding about $25\%$ at $p_T^e= 6\,\mathrm{GeV}$ in the 
unpolarized case shown in the upper panel of Fig.~\ref{fig:figure6}.
Eventually, at somewhat larger values of $p_T^e$,
it becomes dominant, as was shown in \cite{ref:cacciari}.

%
%
%
\begin{figure}[th]
\vspace*{-0.35cm}
\includegraphics[width=0.52\textwidth]{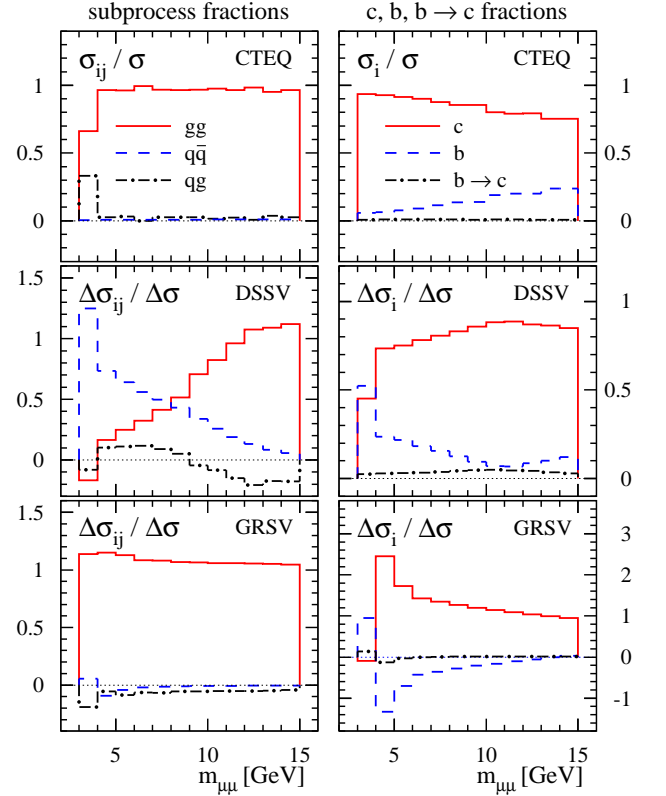}
\vspace*{-0.35cm}
\caption{\label{fig:figure9}
Same as in Fig.~\ref{fig:figure8} but now for the
invariant mass spectrum of muon-muon correlations
shown in Fig.~\ref{fig:figure5}.}
\end{figure}
As for the subprocess fractions, the corresponding results in the
polarized case depend on the choice of parton densities,  
and cancellations and possible sign changes of the individual 
hard scattering subprocesses $d\Delta\hat{\sigma}_{ab}$
further complicate their interpretation. 
The only feature common to both the unpolarized and 
the polarized inclusive electron spectra
is the smallness of the $b\rightarrow c\rightarrow e$ cascade decay 
contribution.
The results obtained with the DSSV set (middle panel) show a very sizable
bottom contribution, exceeding the $c\rightarrow e$ decay 
above $p_T^e\simeq 4\,\mathrm{GeV}$. However, this is due to a sign change
of the $c\rightarrow e$ cross section at $p_T^e\simeq 7\,\mathrm{GeV}$, and
above, $c\rightarrow e$ and $b\rightarrow e$ contribute on equal footing.
Choosing the GRSV(std) distributions instead (lower panel), both the 
$c\rightarrow e$ and the $b\rightarrow e$ cross sections change from
positive to negative at 2 and $4\,\mathrm{GeV}$, respectively, with
$b\rightarrow e$ starting to be the dominant contribution above
$p_T^e\sim 6\,\mathrm{GeV}$.

The fractional contributions of the
different partonic hard scattering processes (left-hand-side) 
and heavy flavor decays (right-hand-side)
to the single-inclusive muon cross section shown
in Figs.~\ref{fig:figure3} are given in Fig.~\ref{fig:figure7}.
The results are qualitatively very similar to the ones depicted
in Fig.~\ref{fig:figure6} and discussed above.
Again, the polarized subprocess fractions very much depend on the choice
of parton densities, and the interpretation is obscured by 
sign changes and large cancellations among the different contributions.
Compared to the single-inclusive electron spectrum at central rapidities,
bottom decays contribute less to the muon transverse momentum spectrum
at $1.2\le|\eta_{\mu}|\le2.2$, even up to $p_T^{\mu}=10\,\mathrm{GeV}$.
Its contribution is rather flat with respect to $p_T^{\mu}$ and amounts to about
$40\%$ ($20\%$) in the (un)polarized case.

%
\begin{figure}[th]
\vspace*{-0.35cm}
\includegraphics[width=0.52\textwidth]{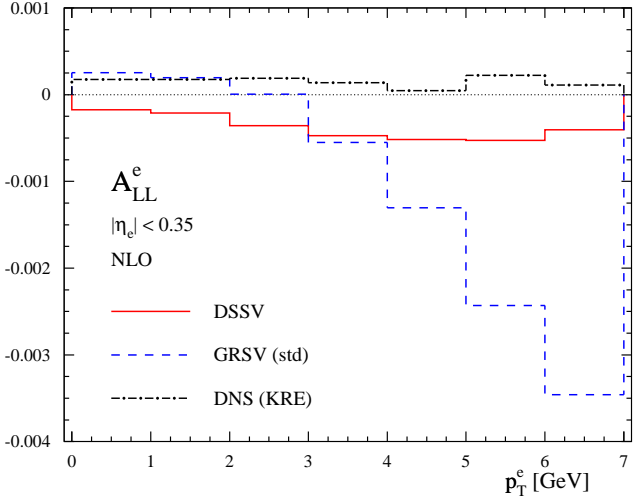}
\vspace*{-0.35cm}
\caption{\label{fig:figure10}
Double-spin asymmetry $A_{LL}^e$ for single-inclusive electrons from
charm and bottom decays at RHIC, computed at NLO accuracy
for three different sets of polarized parton densities:
DSSV \cite{ref:dssv} (solid line), GRSV(std) \cite{ref:grsv} (dashed line), 
and DNS(KRE) \cite{ref:dns} (dot-dashed line).
Electrons are restricted to central rapidities $|\eta_e|\le 0.35$.}
\end{figure}
Corresponding results for the electron-muon and muon-muon invariant
mass distributions, shown in Fig.~\ref{fig:figure4} and \ref{fig:figure5},
can be found in Fig.~\ref{fig:figure8} and \ref{fig:figure9}, respectively.
As before, subprocess fractions can be found on the left-hand-side
and contributions from different heavy flavor decays on the right-hand-side of the plots.
Note that in all panels of Figs.~\ref{fig:figure8} and \ref{fig:figure9},
the bin corresponding to the smallest invariant mass only has a small number
of entries due to the cuts $p_T^{e,\mu}>1\,\mathrm{GeV}$,
and the numerical results for that bin should be taken with caution.
In general, cancellations among different subprocesses are found to be
less pronounced in Figs.~\ref{fig:figure8} and \ref{fig:figure9},
except for small invariant masses, say, below $4\,\mathrm{GeV}$, 
where a sign change in the polarized cross section occurs.

Gluon-gluon fusion is even more dominant for electron-muon and
muon-muon correlations
than for single-inclusive decay lepton observables,
with both $q\bar{q}$ and $qg$ subprocesses 
being negligible in the unpolarized case (upper row).
This is also the case for the polarized cross section if the
GRSV (std) parton distributions are chosen.
For the DSSV set, $q\bar{q}$ annihilation remains
dominant for electron-muon correlations, but gluon-gluon fusion contributes
significantly to muon-muon correlations for $m_{\mu\mu}\gtrsim 10\,\mathrm{GeV}$.
As will be shown in the next Subsection, back-to-back muon-muon correlations 
with $1.2\le|\eta_{\mu}|\le2.2$ probe on average fairly large momentum fractions,
$\langle x\rangle\gtrsim 0.1$, where the DSSV $\Delta g(x)$ is positive and
larger than the sea quark polarizations, such that 
$d\Delta\hat{\sigma}_{gg}>d\Delta\hat{\sigma}_{q\bar{q}}$.
Since the DSSV $\Delta\bar{u}(x)$ turns negative at large $x$, 
there are additional cancellations
among the different quark flavors in the $q\bar{q}$ annihilation channel,
as $\Delta u(x_1)\Delta\bar{u}(x_2)<0$ and $\Delta d(x_1)\Delta\bar{d}(x_2)>0$.

%
\begin{figure}[th]
\vspace*{-0.35cm}
\includegraphics[width=0.52\textwidth]{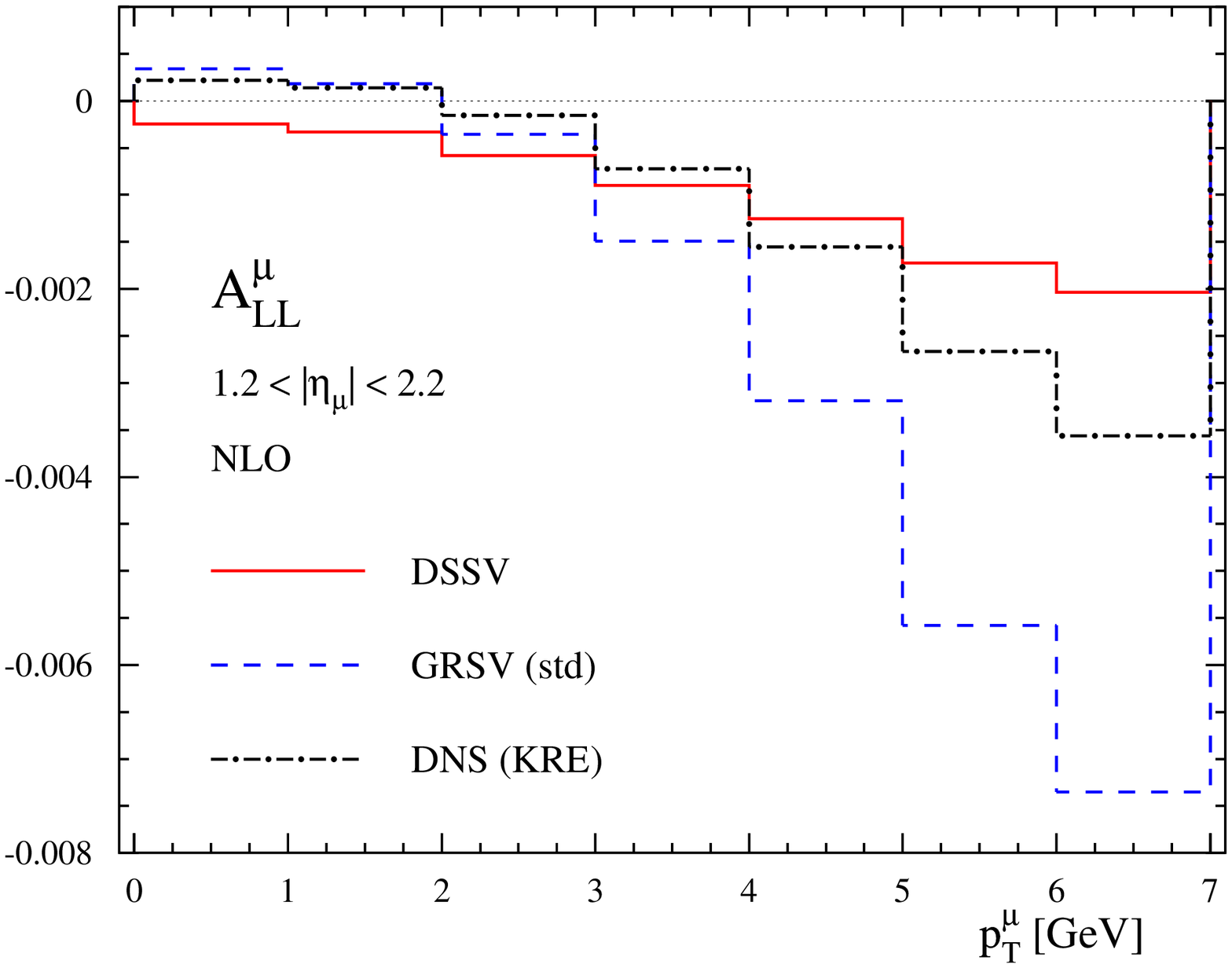}
\vspace*{-0.35cm}
\caption{\label{fig:figure11}
Same as in Fig.~\ref{fig:figure10} but for the
single-inclusive muon spectrum at forward rapidities
$1.2\le|\eta_{\mu}|\le 2.2$.}
%
%
\vspace*{-0.35cm}
\includegraphics[width=0.52\textwidth]{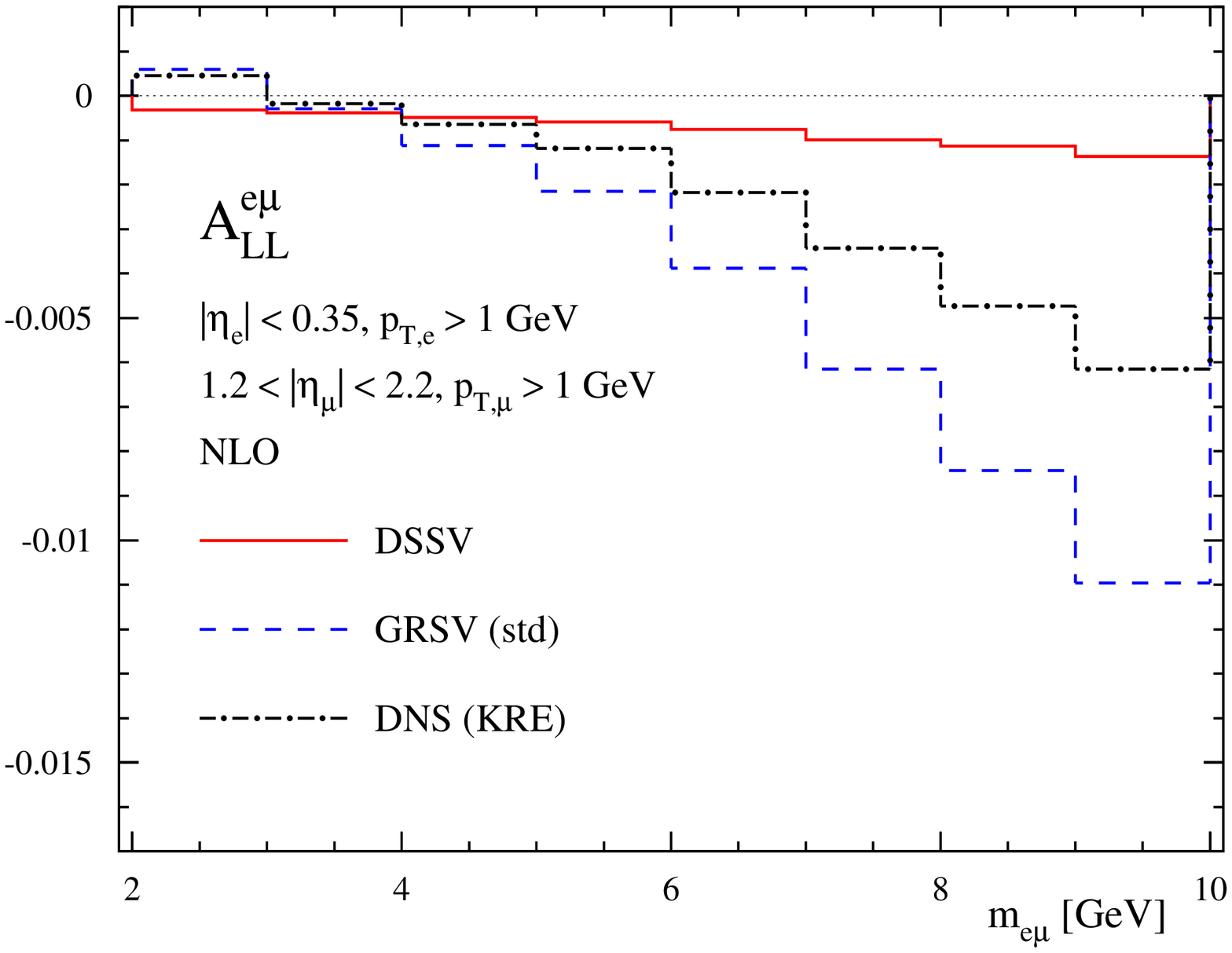}
\vspace*{-0.35cm}
\caption{\label{fig:figure12}
Same as in Fig.~\ref{fig:figure10} but for the
invariant mass spectrum for electron-muon correlations.
Electrons and muons are restricted to
$|\eta_e|\le 0.35$ and $1.2\le|\eta_{\mu}|\le 2.2$,
respectively. In addition, we demand $p_T^e\ge 1\,\mathrm{GeV}$.}
\end{figure}

As in Figs.~\ref{fig:figure6} and \ref{fig:figure7},
the contribution from the cascade $b\rightarrow c\rightarrow e,\mu$ decay
is found to be negligible for lepton-lepton correlations. In particular,
the muons in back-to-back correlations originate mainly from charm decays,
as can be seen on the right-hand-side of Fig.~\ref{fig:figure9}. 
At the same invariant lepton-lepton mass, bottom quark decays contribute 
more significantly to electron-muon correlations than to muon-muon correlations for
$m_{e\mu},m_{\mu\mu}>5\,\mathrm{GeV}$.

\subsection{Double-Spin Asymmetries \label{sec:all}}
%
%
\begin{figure}[th]
\vspace*{-0.35cm}
\includegraphics[width=0.52\textwidth]{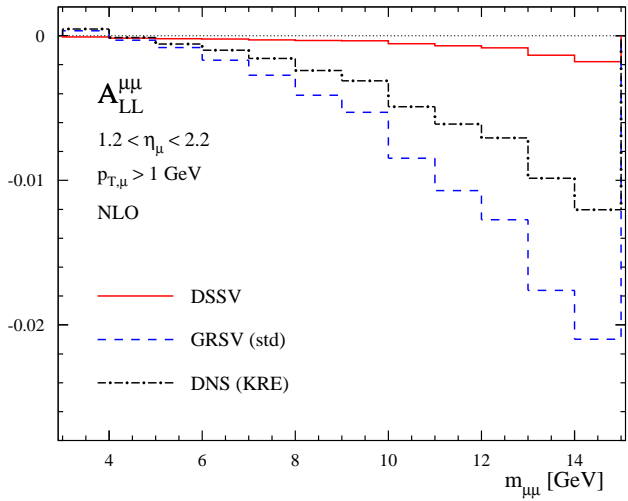}
\vspace*{-0.35cm}
\caption{\label{fig:figure13}
Same as in Fig.~\ref{fig:figure10} but for the
invariant mass spectrum for muon-muon correlations.
The muons are restricted to forward rapidities 
$1.2\le|\eta_{\mu}|\le 2.2$ and have to be in opposite
hemispheres. In addition, we demand $p_T^{\mu}\ge 1\,\mathrm{GeV}$.}
\end{figure}
The quantities of actual interest in experiments exploiting polarized
beams and targets are the double-spin asymmetries $A_{LL}$ 
defined in Eq.~(\ref{eq:all}). Experimental normalization uncertainties
conveniently cancel to a large extent in the ratio (\ref{eq:all}).
In general, this does not happen for higher order QCD corrections or 
the various sources of theoretical ambiguities as we shall demonstrate below.
Nevertheless, it is often erroneously assumed that LO estimates 
for $A_{LL}$ give reliable results which can be used in quantitative QCD
analyses.

We start by giving expectations
for various double-spin asymmetries at NLO accuracy 
in Figs.~\ref{fig:figure10} - \ref{fig:figure13},
based on the polarized and unpolarized cross sections 
for decay lepton transverse momentum and invariant mass distributions
presented in Figs.~\ref{fig:figure2} - \ref{fig:figure5}.
Apart from our default choice of DSSV polarized parton densities \cite{ref:dssv},
which leads to very small asymmetries throughout,
we adopt also two alternative sets, 
GRSV(std)~\cite{ref:grsv} and DNS(KRE)~\cite{ref:dns}.
Unlike DSSV, both sets are characterized by a positive gluon polarization of moderate size and an
almost $SU(3)$ symmetric sea. We refrain from using outdated models with a large,
but strongly disfavored gluon polarization in the $x$ range already 
probed by RHIC $pp$ and fixed target data 
\cite{ref:hermes,ref:smc,ref:compass-2had,ref:compass-charm,ref:phenix,ref:star}.

With the exception of the double-spin asymmetry $A_{LL}^e$ for single-inclusive electrons from
charm and bottom decays shown in Fig.~\ref{fig:figure10},
differences in the results obtained with GRSV(std) and DNS(KRE) parton distributions
are readily explained by the slightly larger $\Delta g(x)$ in the GRSV set.
For $A_{LL}^e$, the result based on the DNS(KRE) set is strongly affected by cancellations
between the $gg$ subprocess on the one hand, and the $q\bar{q}$, $qg$ processes on the
other hand, leading to an essentially zero spin asymmetry in the $p_T^e$ range shown.
Cancellations among the different subprocesses contributing with different sign
are less pronounced for the GRSV set,
see the lower left panel of Fig.~\ref{fig:figure6}, due to a significantly less
negative $\Delta \bar{u}$ density at $x\simeq 0.1$.
Cancellations in conjunction with the smallness of $A_{LL}^e$ for all $p_T^e$
make this observable not really suited for studies of the nucleon's spin structure.

The double-spin asymmetry $A_{LL}^{\mu}$ for single-inclusive decay muons
at $1.2\le|\eta_{\mu}|\le 2.2$, presented in Fig.~\ref{fig:figure11}, shows a much better
correlation of the size of $\Delta g(x)$ and $A_{LL}^{\mu}$. As was demonstrated in
Fig.~\ref{fig:figure7}, cancellations among the different subprocesses are less 
pronounced than for $A_{LL}^e$. For the same value of transverse momentum,
the obtained spin asymmetries are about a factor of two larger for $A_{LL}^{\mu}(p_T^{\mu})$
than for $A_{LL}^e(p_T^e)$. However, they are still significantly smaller than projections
based on very large positive, but outdated gluon polarizations like GRSV(max) \cite{ref:grsv},
see, e.g., Fig.~3 in Ref.~\cite{ref:phenix-polhq}.
Nevertheless, with sufficient statistics accumulated, spin asymmetries of ${\cal{O}}(0.5\%)$
should be measurable. Based on the cross sections given in Figs.~\ref{fig:figure2} and
\ref{fig:figure3}, we estimate that an integrated luminosity of about 
$1\,\mathrm{fb}^{-1}$ is required.

The best suited observables related to heavy flavor hadroproduction in longitudinally
polarized $pp$ collisions at RHIC are double-spin asymmetries for invariant mass spectra 
of electron-muon and muon-muon correlations shown in Fig.~\ref{fig:figure12} and 
\ref{fig:figure13}, respectively.
At small invariant mass, the corresponding cross sections are 
smaller than for single-inclusive transverse momentum distributions 
at similar values of $p_T^{e,\mu}$, but fall off much slower with 
increasing invariant mass, cf.~Figs.~\ref{fig:figure2} - \ref{fig:figure5}. 
This should allow for measurements of $A_{LL}^{e\mu}$ and $A_{LL}^{\mu\mu}$
up to $m_{e\mu,\mu\mu}=10\div15\,\mathrm{GeV}$, where spin asymmetries can be up to the
$1\div2\,\%$ level for the GRSV(std) and DNS(KRE) parton densities.
An integrated luminosity of a few hundred $\mathrm{pb}^{-1}$ should be sufficient.
In addition, the size of $A_{LL}^{e\mu,\mu\mu}$ and $\Delta g(x)$ in the relevant range of
momentum fractions $x$, see below, are nicely correlated.
In the absence of shifted vertex detectors at the RHIC experiments, $A_{LL}^{\mu\mu}$
is the observable with the cleanest sample of charm decays
for all $m_{\mu\mu}$ shown in Fig.~\ref{fig:figure13} and
irrespective of the set of polarized parton densities used in the calculations,
cf.~Fig.~\ref{fig:figure9}. 

%
\begin{figure}[th]
\vspace*{-0.35cm}
\includegraphics[width=0.5\textwidth]{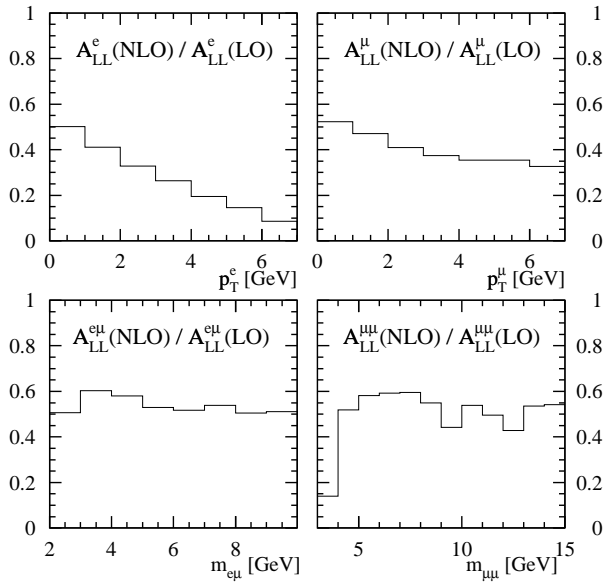}
\vspace*{-0.35cm}
\caption{\label{fig:figure14}
Impact of NLO QCD corrections on the double-spin asymmetries
shown in Figs.~\ref{fig:figure10} - \ref{fig:figure13}.
Depicted is the $K$-factor, i.e., $A_{LL}(NLO)/A_{LL}(LO)$,
computed in each case using the DSSV polarized and 
CTEQ6 unpolarized parton densities.}
\end{figure}
%
%
%
\begin{figure}[th]
\vspace*{-0.35cm}
\includegraphics[width=0.52\textwidth]{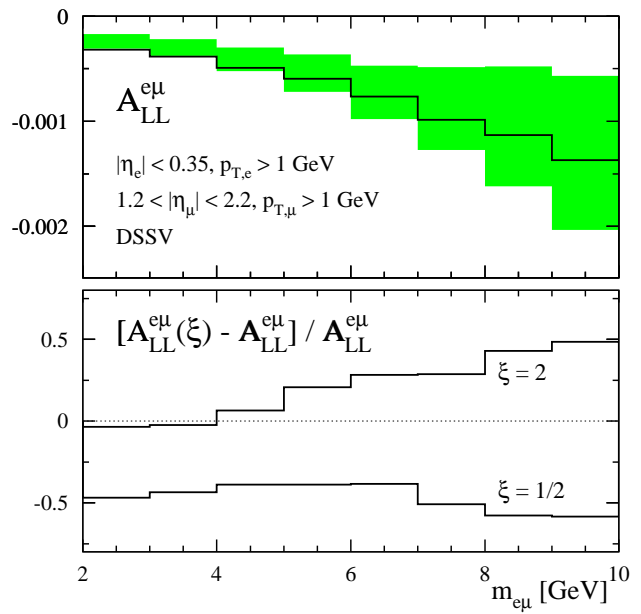}
\vspace*{-0.35cm}
\caption{\label{fig:figure15}
Impact of variations of the scales 
$\mu_f=\mu_r=\xi(m_Q^2+[(p_T^Q)^2+(p_T^{\bar{Q}})^2]/2)^{1/2}$
on the double-spin asymmetry
$A_{LL}^{e\mu}$ for electron-muon correlations at NLO accuracy.
The upper panel shows $A_{LL}^{e\mu}$ computed with the DSSV parton densities
for $\xi=1$ (solid line) and in the range $1/2\le\xi\le2$ (shaded band).
The lower panel gives the relative deviation of $A_{LL}^{e\mu}$
for $\xi=1/2,\,2$ with respect to the result 
obtained for our default value $\xi=1$.}
\end{figure}
Figure~\ref{fig:figure14} clearly illustrates the inadequacy of computing 
double-spin asymmetries based on LO estimates of heavy flavor cross sections.
Depicted is the $K$-factor, i.e., $A_{LL}(NLO)/A_{LL}(LO)$,
for all double-spin asymmetries presented in 
Figs.~\ref{fig:figure10} - \ref{fig:figure13},
computed in each case using the DSSV polarized and 
CTEQ6 unpolarized parton densities. Other sets of polarized parton densities
yield qualitatively very similar results.
On average, LO estimates for $A_{LL}$ are about a factor of two larger than
corresponding calculations at NLO accuracy and depend, in case of
the single-inclusive observables $A_{LL}^e$ and  $A_{LL}^{\mu}$, strongly on 
$p_T^{e,\mu}$.
This reflects the difference of $K$-factors 
for the polarized and unpolarized cross sections found in 
Figs.~\ref{fig:figure2} - \ref{fig:figure5}
and invalidates any approximation based on constant $K$-factors or
the idea that higher order QCD corrections cancel in $A_{LL}$.

As was already illustrated in Fig.~\ref{fig:figure5b},
theoretical uncertainties associated with the actual choice of the parameters
$\alpha_{c,b}$ in the non-perturbative function $D^{Q\rightarrow H_Q}(z)$
given in Eq.~(\ref{eq:kart}) cancel to a large extent in 
double-spin asymmetries.
Unfortunately, this is not the case for ambiguities 
related to the choice of scales $\mu_{f,r}$.
As we have discussed in Sec.~\ref{sec:scale}, the dependence
of unpolarized heavy flavor cross sections on variations of $\mu_{f,r}$ is in general
more pronounced than in the polarized case, see, e.g., Fig.~\ref{fig:figure4}
for electron-muon correlations. This can cause sizable ambiguities also for
ratios of cross sections, like double-spin asymmetries.
As a representative example, we show in Fig.~\ref{fig:figure15} the
dependence of the double-spin asymmetry $A_{LL}^{e\mu}$ for electron-muon
correlations on variations of  $\mu_{f,r}$.
The shaded band in the upper panel of Fig.~\ref{fig:figure15} illustrates
the uncertainty on $A_{LL}^{e\mu}$ if
$\mu_f=\mu_r=\xi(m_Q^2+[(p_T^Q)^2+(p_T^{\bar{Q}})^2]/2)^{1/2}$ 
are varied simultaneously in the range $1/2\le\xi\le2$.
The lower panel gives the relative deviation of $A_{LL}^{e\mu}$
for $\xi=1/2,\,2$ with respect to the result 
obtained for our default value $\xi=1$.

The scale uncertainties are quite substantial and not uniform 
as a function of the invariant mass $m_{e\mu}$.
Nevertheless, the asymmetries obtained with the DSSV parton densities 
are still much smaller than for sets 
with larger gluon polarizations, like DNS(KRE) or GRSV(std),
as can be inferred by comparing with the results given in Fig.~\ref{fig:figure12}.
Qualitatively similar effects as in  Fig.~\ref{fig:figure15} are found
for the other double-spin asymmetries discussed in this Subsection.
We refrain from varying $\mu_f$ and $\mu_r$ independently, which increases
the uncertainties only slightly for the specific observables and kinematics
we are interested in, cf.~Fig.~\ref{fig:figure5a}.
Similar observations apply to variations of the heavy quark masses $m_{c,b}$, which,
in principle, need to be considered as well, see~Fig.~\ref{fig:figure5a}.

\begin{figure}[bth]
\vspace*{-0.35cm}
\includegraphics[width=0.5\textwidth]{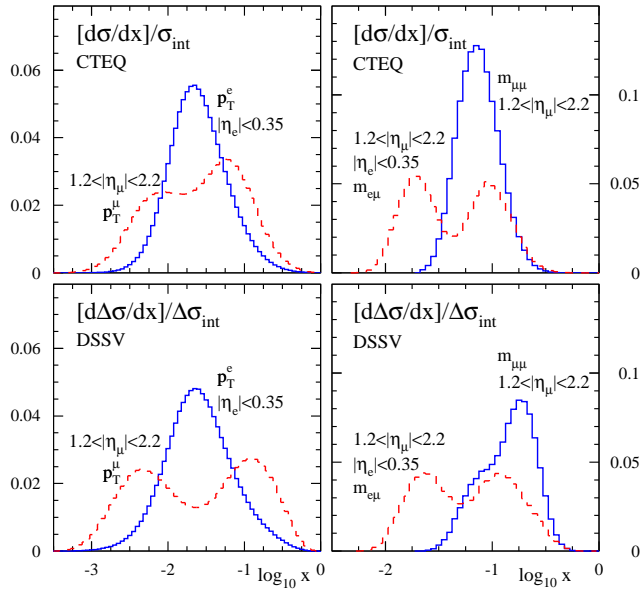}
\vspace*{-0.45cm}
\caption{\label{fig:figure16}
Typical range of momentum fractions $x$ at which
the parton densities in Eq.~(\ref{eq:xsec-fact}) are probed for
the integrated unpolarized (upper row) and polarized
(lower row) cross sections for leptons from heavy flavor
decays at RHIC shown in Figs.~\ref{fig:figure2}-\ref{fig:figure5}.
The left panels give the results for the single-inclusive
electron (solid lines) and muon (dashed lines) cross sections.
Results for the electron-muon (dashed lines)
and muon-muon (solid lines) correlations are presented in the right panels.}
\end{figure}

Finally, we estimate the range of momentum fractions $x$ at which the
(un)polarized parton densities in  Eq.~(\ref{eq:xsec-fact}) 
are predominantly probed for the different single-inclusive 
and lepton-lepton correlation observables discussed in this paper.
Figure~\ref{fig:figure16} shows the corresponding
cross sections differential in $x$.
In each case, we have integrated over the angular acceptance for
detecting electrons and/or muons with the PHENIX experiment at RHIC
as well as over all transverse momenta of the decay leptons. 
As before, an additional cut $p_T^{e,\mu}>1\,\mathrm{GeV}$ is imposed
for lepton-lepton correlations.
All $x$ distributions are normalized to the respective integrated cross section
$(\Delta)\sigma_{int}$.
%
%

Since small transverse momenta probe also the smallest possible $x$ values,
Fig.~\ref{fig:figure16} gives a rough idea of the lowest possible 
momentum fractions accessible in heavy quark hadroproduction at RHIC with
a c.m.s.~energy of $\sqrt{S}=200\,\mathrm{GeV}$.
As can be seen, single-inclusive observables, shown in the panels on the left,
receive contributions from $x$ values as low as $10^{-3}$, but the
majority of events has $\langle x\rangle\simeq 0.01$.
Both, electron-muon and muon-muon correlations, displayed on the
right-hand-side of Fig.~\ref{fig:figure16}, probe on average 
larger values of $x$. In particular, the latter observable can be used to study
parton densities at $\langle x\rangle\simeq 0.1$ or higher.
The entire suite of possible observables related to heavy flavor hadroproduction
at RHIC can cover a wide range in $x$ and has the potential to provide novel information
on the spin structure of the nucleon and the applicability of perturbative QCD in
polarized hard scattering processes.

\section{Summary and Outlook}
%
We have presented a flexible parton-level Monte Carlo program to
compute heavy flavor distributions and correlations at NLO accuracy 
in longitudinally polarized $pp$ collisions. Experimental acceptance cuts,
the hadronization of the produced heavy quark pair into $D$ and $B$ mesons, 
and their subsequent semi-leptonic decays can be included in phenomenological
applications.

Heavy flavor hadroproduction receives its importance for the field of spin physics 
from its partonic hard scattering processes, which differ from their
counterparts for light hadron and jet production utilized in global QCD analyses
so far. Upcoming measurements of heavy flavor production at RHIC 
will further our current understanding of factorization in the presence of spin
and test the notion of universality for polarized parton densities.

We have performed a comprehensive phenomenological study of various 
observables where heavy quarks are identified through their semi-leptonic decays
into electrons or muons. 
Such measurements can be done once sufficient statistics has been collected 
in polarized proton-proton collisions at RHIC and do
not require the completion of vertex detector upgrades.
Decay lepton correlations turned out to be particularly suited probes for
the spin structure of the nucleon, and experimentally relevant 
double-spin asymmetries of about one percent can be expected even for present-day 
gluon polarizations of rather moderate size.
Contrary to naive expectations based on unpolarized results, gluon-gluon
fusion is not necessarily the dominant channel for heavy flavor production
in longitudinally proton-proton collisions at RHIC.

Next-to-leading order QCD corrections are in general more pronounced for 
unpolarized than for polarized heavy flavor production cross sections, such
that they do not cancel in the ratio defining double-spin asymmetries.
Also, theoretical uncertainties estimated by varying factorization and
renormalization scales are usually smaller for spin-dependent observables.

The technical methods and their implementation into a parton-level Monte Carlo program can
be straightforwardly extended to deal also with the spin-dependent photoproduction
of heavy quarks. This will allow one to analyze already existing data
for charm production consistently at NLO accuracy within future global QCD analyses
of polarized parton densities. 
In addition, one can assess the physics impact of heavy flavor distributions and correlations
obtained at a possible future polarized lepton-nucleon collider like the EIC.

\begin{acknowledgments}
We thank Ming Xiong Liu for useful discussions and Ramona Vogt for
providing us with the electron spectra used in Ref.~\cite{ref:cacciari}.
J.R.\ is supported by a grant of the ``Cusanuswerk'', Bonn,
Germany.
M.S.\ acknowledges partial support of the Initiative and
Networking Fund of the Helmholtz Association, contract HA-101
("Physics at the Terascale").
This work was supported in part by the ``Bundesministerium f\"ur
Bildung und Forschung'' (BMBF), Germany.
\end{acknowledgments}

\section*{Appendix}
%
In the Appendix we collect for completeness some additional
details of the calculation and some explicit expressions
which were omitted in \cite{ref:nlo-pol} but might be
useful for the reader.

First of all, we recall the LO partonic cross sections for open
heavy flavor hadroproduction, which are needed for the factorization
of collinear singularities. They also emerge in the soft gluon limit
of the NLO $2\rightarrow 3$ real emission contributions.
The spin-dependent, color-averaged matrix elements squared for the tree-level processes
in (\ref{eq:born-proc}) read in $d=4-2\varepsilon$ dimensions:
\begin{eqnarray}
\label{eq:mgglo}
\nonumber
\Delta|M_{gg}|^2 &=& \frac{1}{2s}  (4\pi\alpha_s)^2
\frac{1}{2(N_C^2-1)}\\
&\times& \left[2 C_F-C_A \frac{2 t_1 u_1}{s^2}\right] 
\Delta B_{QED}\,,\\
\label{eq:mqqlo}
\Delta|M_{q\bar{q}}|^2 &=& \frac{1}{2s} (4\pi\alpha_s)^2 \frac{C_F}{N_C} \Delta A_{QED}\,,
\end{eqnarray}
where
\begin{eqnarray}
\label{eq:bqed}
\Delta B_{QED} &=&\left(\frac{t_1}{u_1}+\frac{u_1}{t_1}\right)
\left(\frac{2 m_Q^2 s}{t_1 u_1}-1\right)\,,\\
\label{eq:aqed}
\Delta A_{QED} &=& -\frac{t_1^2+u_1^2}{s^2}-\frac{2 m_Q^2}{s}-\varepsilon\,.
\end{eqnarray}
Here, $N_C$ denotes the number of colors, 
$C_A=N_C$, and $C_F=(N_C^2-1)/(2N_C)$.
Contrary to the unpolarized case \cite{ref:nlo-unpol,ref:nlo-unpol2},
$\Delta B_{QED}$ receives no ${\cal{O}}(\varepsilon)$ contributions.
The Mandelstam variables used in Eqs.~(\ref{eq:mgglo}) - (\ref{eq:aqed}) 
are defined by
\begin{eqnarray}
\label{eq:mandel}
\nonumber
s &=&(p_1+p_2)^2\,,\\
\nonumber
t_1&=&(p_1-k_1)^2-m_Q^2\,,\\
u_1&=&(p_1-k_2)^2-m_Q^2\,,
\end{eqnarray}
where $s+t_1+u_1=0$.
$p_{1,2}$ are the momenta of the incoming partons, and
$k_{1}$ and $k_2$ are the momenta of the produced heavy quark
and antiquark, respectively.
Together with the appropriate 
two-body phase-space factor $d{\mathrm{PS}}_2$
in $d$ dimensions \cite{ref:nlo-unpol},
\begin{eqnarray}
\label{eq:dps2}
\nonumber
d\mathrm{PS}_2 &=& \frac{2\pi}{s} \left[(4\pi)^{2-\varepsilon}
\Gamma(1-\varepsilon)\right]^{-1}
\left(\frac{t_1 u_1-m_Q^2 s}{s}\right)^{-\varepsilon}\\
&\times& \delta(s+t_1+u_1)dt_1 du_1\,,
\end{eqnarray}
the $gg$ and $q\bar{q}$ Born cross sections can be written as
\begin{equation}
\label{eq:born-xsec}
\frac{d^2\Delta\hat{\sigma}_{ab}}{dt_1du_1} =
F_{\varepsilon}\, \Delta|{M}_{ab}|^2\, \delta (s+t_1+u_1)\,,
\end{equation}
where
\begin{equation}
\label{eq:feps}
F_{\varepsilon} \equiv \frac{\pi}{s^2}\left[(4\pi)^{2-\varepsilon}
\Gamma(1-\varepsilon)\right]^{-1}
\left(\frac{t_1 u_1-m^2 s}{\mu^2 s}\right)^{-\varepsilon}
\end{equation}
collects all phase-space factors given in
Eq.~(\ref{eq:dps2}) and the flux factor $1/(2s)$ included in
Eqs.~(\ref{eq:mgglo}) and (\ref{eq:mqqlo}).
The mass parameter $\mu$ is introduced to keep the strong coupling
dimensionless in $d$ dimensions.
In the limit $\varepsilon\rightarrow 0$, $F_{\varepsilon}$
reduces to $1/(16 \pi s^2)$.
In the c.m.s.~frame of the incoming partons, $d{\mathrm{PS}}_2$
and $d\Delta\hat{\sigma}_{ab}$ can be conveniently expressed 
in terms of the scattering angle between $\vec{p}_1$ and
$\vec{k}_1$ by using the relation
\begin{equation}
\label{eq:t1-cos}
t_1=-\frac{s}{2}\left(1-\beta \cos\theta_1 \right)\,,
\end{equation}
where $\beta^2=1-4m_Q^2/s=1-\rho$.
The corresponding unpolarized Born cross sections in $d$ dimensions can be found
in Ref.~\cite{ref:nlo-unpol}.

Next, we give explicit expressions for the 
$\tilde{\rho}$ and $\omega$ prescriptions introduced 
in Eq.~(\ref{eq:xy-expand}) to
regularize soft and collinear regions of phase-space.
The distributions are defined as follows \cite{ref:mnr}:
\begin{align}
\label{eq:distrib}
\nonumber
&\int_{\tilde{\rho}}^1h(x)\left(\frac{1}{1-x}\right)_{\tilde{\rho}} dx
=\int_{\tilde{\rho}}^1\frac{h(x)-h(1)}{1-x} dx,\\
\nonumber
&\int_{\tilde{\rho}}^1h(x)\left(\frac{\log(1-x)}{1-x}\right)_{\tilde{\rho}} dx\\
\nonumber
&\quad\quad\quad\quad=\int_{\tilde{\rho}}^1[h(x)-h(1)]\frac{\log(1-x)}{1-x} dx,\\
\nonumber
&\int_{1-w}^1h(y)\left(\frac{1}{1-y}\right)_{w} dy
=\int_{1-w}^1\frac{h(y)-h(1)}{1-y} dy,\\
\nonumber
&\int_{-1}^{-1+w}h(y)\left(\frac{1}{1+y}\right)_{w} dy\\
&\quad\quad\quad\quad=\int_{-1}^{-1+w}\frac{h(y)-h(-1)}{1+y} dy.
\end{align}
$h$ denotes an arbitrary ``test function'', which is sufficiently regular
in the limits $x\rightarrow 1$ and $y\rightarrow \pm 1$.
In a numerical implementation, 
the distributions (\ref{eq:distrib}) need to be used 
only if $x$ and $y$ are sampled in the ranges
$\tilde{\rho}<x<1$ and $1-\omega<y<1$ or $-1<y<-1+\omega$,
respectively, in the Monte Carlo integration.

Following Ref.~\cite{ref:nlo-unpol},
the soft, $x\rightarrow 1$ limit of the $2\rightarrow 3$ matrix elements squared 
is derived by applying the eikonal approximation, i.e.,
by taking the limit when the momentum of the emitted gluon gets
soft. The obtained soft matrix elements squared for $q\bar{q}$ and $gg$
scattering have the same form as in the unpolarized case 
\cite{ref:nlo-unpol} but with the Born results
replaced by their polarized counterparts $\Delta A_{QED}$ and
$\Delta B_{QED}$, given in Eq.~(\ref{eq:aqed}) and (\ref{eq:bqed}),
respectively. 
The expressions can be straightforwardly 
integrated over $x$ and $y$ in (\ref{eq:dps3}), and we obtain
for $d\Delta\hat{\sigma}_{q\bar{q}}^{(s)}$
\begin{widetext}
\begin{align}
\label{eq:soft-qqbf}
\frac{d\Delta\hat{\sigma}_{q\bar{q},F}^{(s)}}{dt_1du_1}&=
2 \frac{C_F^2}{N_C}F_\varepsilon G_\varepsilon \alpha_s^3 \tilde{\beta}^{-4\varepsilon} 
\Delta A_{QED} 
\Bigg\{\frac{2}{\varepsilon^2}+\frac{2}{\varepsilon}-\frac{2}{\varepsilon}
\ln\frac{sm^2_Q}{t_1u_1}-\frac{8}{\varepsilon}\ln\frac{t_1}{u_1}+2\
-8\ln\varkappa\ln\frac{t_1}{u_1}+\ln^2\frac{sm^2_Q}{t_1u_1}\notag\\
&\qquad{}
+2\mathrm{Li}_2\left(1-\frac{sm^2_Q}{t_1u_1}\right)+8\mathrm{Li}_2\left(1-
\frac{t_1}{\varkappa u_1}\right)
-8\mathrm{Li}_2\left(1-\frac{u_1}{\varkappa t_1}\right)
+\frac{2(2m^2_Q-s)}{s\beta}\left[\ln\varkappa-S(\varkappa)\right]
\Bigg\}\delta(s+t_1+u_1)\\
\label{eq:soft-qqba}
\frac{d\Delta\hat{\sigma}_{q\bar{q},A}^{(s)}}{dt_1du_1}&=
\frac{C_A C_F}{N_C} F_\varepsilon G_\varepsilon \alpha_s^3 \tilde{\beta}^{-4\varepsilon} 
\Delta A_{QED}
\Bigg\{\frac{2}{\varepsilon}\ln\frac{sm^2_Q}{t_1u_1}
+\frac{6}{\varepsilon}\ln\frac{t_1}{u_1}-\ln^2\frac{sm^2_Q}{t_1u_1}
+6\ln\varkappa\ln\frac{t_1}{u_1}
+\ln^2\frac{t_1}{u_1}-\ln^2\varkappa \notag\\
&\qquad{}
-2\mathrm{Li}_2\left(1-\frac{sm^2_Q}{t_1u_1}\right)-6\mathrm{Li}_2\left(1-
\frac{t_1}{\varkappa u_1}\right) 
+6\mathrm{Li}_2\left(1-\frac{u_1}{\varkappa t_1}\right)
+\frac{2(2m^2_Q-s)}{s\beta}S(\varkappa)
\Bigg\}\delta(s+t_1+u_1)\,,
\end{align}
\end{widetext}
where we have used
\begin{align}
\nonumber
\label{eq:skappa}
S(\varkappa) &= \frac{1}{\varepsilon} \ln \varkappa + 
2\ln \varkappa \ln(1-\varkappa^2) - \ln^2\varkappa\\
&+\,\, \mathrm{Li}_2 \varkappa^2 -\zeta(2)\,,\\
\label{eq:gepsilon}
G_\varepsilon &= 64\pi 
e^{-\varepsilon[\gamma_E-\ln(4\pi)]} 
(1-\frac{3}{2}\zeta(2)\varepsilon^2)
\left(\frac{m_Q^2}{\mu^2}\right)^{-\varepsilon}\,,
\end{align}
and $\varkappa\equiv(1-\beta)/(1+\beta)$.
The dilogarithm function $\mathrm{Li}_2(\varkappa)$ is defined
as in Ref.~\cite{ref:dilog}, and $\zeta(2)=\pi^2/6$ denotes the
Riemann Zeta function. $G_\varepsilon$ in (\ref{eq:gepsilon})
parameterizes the difference of the $2\rightarrow 3$ 
and $2\rightarrow 2$ phase-space factors, the latter
given by $F_\varepsilon$ in Eq.~(\ref{eq:feps}).
As in Ref.~\cite{ref:nlo-unpol}, we have split up the result
for $d\Delta\hat{\sigma}_{q\bar{q}}^{(s)}$ into contributions
from different color structures.
The results for $d\Delta\hat{\sigma}_{q\bar{q},F}^{(s)}$
and  $d\Delta\hat{\sigma}_{q\bar{q},A}^{(s)}$ in
Eq.~(\ref{eq:soft-qqbf}) and (\ref{eq:soft-qqba}), respectively,
agree with the corresponding unpolarized expressions in Ref.~\cite{ref:nlo-unpol}
after replacing $\Delta A_{QED}$ by $A_{QED}$.
With the help of (\ref{eq:t1-cos}), 
$d\Delta\hat{\sigma}_{q\bar{q}}^{(s)}/dt_1du_1$ can be easily
transformed into $d\Delta\hat{\sigma}_{q\bar{q}}^{(s)}/d\cos\theta_1$
used in Eq.~(\ref{eq:xsec-soft}).

Likewise, we obtain for the different color factors
contributing to $d\Delta\hat{\sigma}_{gg}^{(s)}$:
\begin{widetext}
\begin{align}
\label{eq:soft-ggqed}
\frac{d\Delta\hat{\sigma}^{(s)}_{gg,QED}}{dt_1du_1}
&=\frac{4 C_F^2}{N_C^2-1}
F_\varepsilon G_\varepsilon \alpha_s^3 \tilde{\beta}^{-4\varepsilon} 
\Delta B_{QED}
\left\{\frac{1}{\varepsilon}+1+\frac{2 m^2_Q-s}{s\beta}
\left[\ln\varkappa-S(\varkappa)\right]\right\}\delta(s+t_1+u_1)\,,\\
\label{eq:soft-gga}
\frac{d\Delta\hat{\sigma}^{(s)}_{gg,A}}{dt_1du_1}
&= \frac{C_A^2}{N_C^2-1} F_\varepsilon G_\varepsilon \alpha_s^3
\tilde{\beta}^{-4\varepsilon} 
\Delta B_{QED}
\Bigg\{
-\frac{2t_1u_1}{s^2}\left(\frac{1}{\varepsilon}+1\right)
+ \left(1-\frac{2t_1u_1}{s^2}\right)\notag\\
&\quad\times \left[\frac{2}{\varepsilon^2}
-\frac{1}{\varepsilon}\ln\frac{m^2_Qs}{t_1u_1}
+\frac{1}{2}
   (\ln^2\frac{m^2_Qs}{t_1u_1}+\ln^2\frac{t_1}{u_1}-\ln^2\varkappa)
+\mathrm{Li}_2\left(1-\frac{m^2_Qs}{t_1u_1}\right)\right]\notag\\
&\quad{}+\frac{t_1^2-u_1^2}{s^2}\left[
  \ln\frac{t_1}{u_1}\left(-\frac{2}{e}+\ln\varkappa\right)
  -\mathrm{Li}_2\left(1-\frac{t_1}{u_1\varkappa}\right)
  +\mathrm{Li}_2\left(1-\frac{u_1}{t_1\varkappa}\right)\right] \notag\\
&\quad{}-\frac{2m^2_Q-s}{s\beta}\left[\frac{2t_1u_1}{s^2}\ln\varkappa-S(\varkappa)\right]
\Bigg\}\delta(s+t_1+u_1)\,,\\
\label{eq:soft-gg1}
\frac{d\Delta\hat{\sigma}^{(s)}_{gg,1}}{dt_1du_1}
&= \frac{1}{N_C^2-1} F_\varepsilon G_\varepsilon \alpha_s^3
\tilde{\beta}^{-4\varepsilon} 
\Delta B_{QED}
\Bigg\{
-\frac{2}{\varepsilon^2}
+\frac{2t_1u_1}{s^2}\left(\frac{1}{\varepsilon}+1\right)
-\ln^2\frac{t_1}{u_1}+\ln^2\varkappa \notag\\
&\quad{}-\frac{2(2m^2_Q-s)}{s\beta}
\left[-\frac{t_1u_1}{s^2}\ln\varkappa
  +\left(1+\frac{t_1u_1}{s^2}\right)S(\varkappa)\right]
\Bigg\}\delta(s+t_1+u_1)\,.
\end{align}
\end{widetext}
Note that we have chosen a slightly different way of organizing the
above results according to their color structure than in Ref.~\cite{ref:nlo-unpol},
but the sum of Eqs.~(\ref{eq:soft-ggqed})-(\ref{eq:soft-gg1}) agrees with
their expression after replacing $\Delta B_{QED}$ by 
its unpolarized counterpart $B_{QED}$ \cite{ref:nlo-unpol}.
The expressions for the virtual corrections $d\Delta\hat{\sigma}^{(v)}_{ab}$
and the finite contributions $d\Delta\hat{\sigma}^{(f)}_{ab}$ in (\ref{eq:xsec-finite})
are too long to be presented here. They are available upon request.

Turning to the collinear, $y\rightarrow \pm 1$ limit of the 
$2\rightarrow 3$ processes in (\ref{eq:nlo-proc}),
we give explicit expressions for the functions $\Delta f_{ab}^{(c\pm)}$ 
appearing in Eq.~(\ref{eq:collinear}).
They read
\begin{align}
\label{eq:coll1}
\Delta f_{gg}^{(c+)}(x,\theta_1) 
&= 32\pi \alpha_s s(1-x)\notag\\ 
&\quad\times\Delta|M_{gg}|^2\left|_{p_1\rightarrow xp_1} \right. \Delta P_{gg}(x)\,,\\
\Delta f_{gg}^{(c-)}(x,\theta_1) 
&= 32\pi \alpha_s s(1-x)\notag\\  
&\quad\times\Delta|M_{gg}|^2\left|_{p_2\rightarrow xp_2} \right. \Delta P_{gg}(x)\,,\\
\Delta f_{qg}^{(c+)}(x,\theta_1) 
&= 32\pi \alpha_s s(1-x)\notag\\    
&\quad\times\Delta|M_{gg}|^2\left|_{p_1\rightarrow xp_1}\right. \Delta P_{gq}(x)\,,\\
\Delta f_{qg}^{(c-)}(x,\theta_1) 
&= 32\pi \alpha_s s(1-x)\notag\\
&\quad\times\Delta|M_{q\bar{q}}|^2\left|_{p_2\rightarrow xp_2}\right. \Delta P_{qg}(x) \,,\\
\Delta f_{q\bar{q}}^{(c+)}(x,\theta_1) 
&= 32\pi \alpha_s s(1-x)\notag\\
&\quad\times\Delta|M_{q\bar{q}}|^2\left|_{p_1\rightarrow xp_1} \right. \Delta P_{qq}(x)\,,\\
\label{eq:coll6}
\Delta f_{q\bar{q}}^{(c-)}(x,\theta_1) 
&= 32\pi \alpha_s s(1-x)\notag\\
&\quad\times\Delta|M_{q\bar{q}}|^2\left|_{p_2\rightarrow xp_2} \right. \Delta P_{qq}(x)\,,
\end{align}
where the Born matrix elements squared in Eqs.~(\ref{eq:mgglo}) and
(\ref{eq:mqqlo}) are to be evaluated with ``shifted'' kinematics.
This is due to the
collinear emission off one of the incoming partons such that only
a fraction $x$ of their original momenta $p_{1,2}$ is available in the subsequent
hard scattering.
The $d=4-2\varepsilon$ dimensional LO polarized splitting functions
$\Delta P_{ij}$ in Eqs.~(\ref{eq:coll1})-(\ref{eq:coll6})
can be found in \cite{ref:nlo-split} and read: 
\begin{eqnarray}
\label{eq:pol-split}
\nonumber
\Delta P_{qq}(x)&=&C_F \Big[ \frac{2}{(1-x)_{\tilde{\rho}}} - 1-x + 3\varepsilon(1-x)\\
\nonumber
           &+& \delta(1-x) \left(\frac{3+\varepsilon}{2}+2\log\tilde{\beta}\right)\Big],\\
\nonumber
\Delta P_{qg}(x)&=&\frac{1}{2}\left[2x-1-2\varepsilon(1-x)\right],\\
\nonumber
\Delta P_{gq}(x)&=&C_F\left[2-x+2\varepsilon(1-x)\right],\\
\nonumber
\Delta P_{gg}(x)&=&2 C_A \Big[\frac{1}{(1-x)_{\tilde{\rho}}} - 2x+1+2\varepsilon(1-x)\\
&+& \delta(1-x)\left(\frac{\beta_0}{2}+\varepsilon\frac{C_A}{6}+2\log\tilde{\beta}\right)\Big]\,,
\end{eqnarray}
with $\beta_0=11C_A/3-2n_f/3$, $n_f$ as the number of active flavors, and where
we have expressed the standard  $1/(1-x)_+$ distributions 
in $\Delta P_{qq}$ and $\Delta P_{gg}$ by the corresponding 
$\tilde{\rho}$-prescriptions defined in (\ref{eq:distrib}).
This amounts to introducing an additional $\log\tilde\beta$ term in the soft $\delta(1-x)$ 
parts of $\Delta P_{qq}$ and $\Delta P_{gg}$ in (\ref{eq:pol-split}).
Note that in Eqs.~(\ref{eq:coll1})-(\ref{eq:coll6}), contributions proportional to
$\delta(1-x)$ do not show up as they are already included as $1/\varepsilon^2$ poles 
in the soft cross sections listed in Eqs.~(\ref{eq:soft-qqbf})-(\ref{eq:soft-gg1}).
In the factorization counter term $d\Delta\sigma_{ab}^{\tilde{c}}$ in (\ref{eq:xsec-counter}),
only four-dimensional splitting functions are needed, i.e., $\varepsilon\rightarrow 0$
in (\ref{eq:pol-split}).


%
\end{document}